\documentstyle[11pt,aaspp4]{article}  
\newcommand{\kms}{\mbox{ km~s$^{-1}$}}

\newcommand{\etal}{\mbox{ et~al.}}

\newcommand{\Nuzc}{\mbox{19,369}}
\newcommand{\Nmul}{\mbox{1,807}}
\newcommand{\Nzwi}{\mbox{18,901}}
\def\gtorder{\mathrel{\raise.3ex\hbox{$>$}\mkern-14mu\lower0.6ex\hbox{$\sim$}}}
\def\ltorder{\mathrel{\raise.3ex\hbox{$<$}\mkern-14mu\lower0.6ex\hbox{$\sim$}}}
\def\pdeg{\ifmmode $\setbox0=\hbox{$^{\circ}$}\rlap{\hskip.11\wd0 .}$^{\circ}}

\begin{document}
\title{The Updated Zwicky Catalog (UZC)
\footnote{Dedicated to the memory of Jim Peters, whose 
friendship, skill and dedication were essential to this work.}
\footnote{This research made use of the NASA/IPAC Extragalactic Database (NED)
  which is operated by the Jet Propulsion Laboratory, Caltech, under contract
  with the National Aeronautics and Space Administration}
\footnote{We have made use in part of finder chart(s)
         obtained using the Guide Stars Selection System Astrometric Support
         Program developed at the Space Telescope Science Institute (STScI is
         operated by the Association of Universities for Research in Astronomy,
         Inc., for NASA) }
}
\author{Emilio E. Falco, Michael J. Kurtz, Margaret J. Geller, 
John P. Huchra, James Peters, Perry Berlind, Douglas J. Mink, 
Susan P. Tokarz and Barbara Elwell}
\affil{Harvard-Smithsonian Center for Astrophysics\\
60 Garden Street, Cambridge MA 02138}

\begin{abstract}

The Zwicky Catalog of galaxies (ZC), with $m_{Zw} \leq 15.5$, has been
the basis for the Center for Astrophysics (CfA) redshift surveys.  To
date, analyses of the ZC and redshift surveys based on it have relied
on heterogeneous sets of galaxy coordinates and redshifts.  Here we
correct some of the inadequacies of previous catalogs by providing:
(1) coordinates with $\lesssim 2\arcsec$ errors for all of the \Nuzc\ 
catalog galaxies, (2) homogeneously estimated redshifts for the
majority (98\%) of the data taken at the CfA (14,632 spectra), and (3)
an estimate of the remaining ``blunder'' rate for both the CfA
redshifts and for those compiled from the literature. For the
reanalyzed CfA data we include a calibrated, uniformly determined
error and an indication of the presence of emission lines in each
spectrum.
We provide redshifts for 7,257 galaxies in the CfA2 redshift survey
not previously published; for another 5,625 CfA redshifts we list the
remeasured or uniformly re-reduced value. Among our new measurements,
\Nmul\ are members of UZC ``multiplets'' associated with the original
Zwicky catalog position in the coordinate range where the catalog is
98\% complete.  These multiplets provide new candidates for
examination of tidal interactions among galaxies.  All of the new
redshifts correspond to UZC galaxies with properties recorded in the
CfA redshift compilation known as ZCAT.  About 1,000 of our new
measurements were motivated either by inadequate signal-to-noise in
the original spectrum or by an ambiguous identification of the galaxy
associated with a ZCAT redshift.
The redshift catalog we include here is $\sim 96$\% complete 
to $m_{Zw} \leq 15.5$, and $\sim 98$\% complete (12,925 galaxies
out of a total of 13,150)
for the right ascension ranges $20^h \leq \alpha_{1950} \leq 4^h$ and
$8^h \leq \alpha_{1950} \leq 17^h$ and declination range $-2\pdeg5 \leq
\delta_{1950} \leq 50^\circ$.  This more complete region includes all of
the CfA2 survey as analyzed to date. The Great Wall structure persists
throughout the Northern survey region.

\end{abstract}


\section{Introduction}

During the last 15 years, wide-angle redshift surveys of the nearby
universe have provided a basis for statistical characterization of the
local large-scale structure of the universe. The CfA (Huchra, Vogeley
\& Geller 1998; Huchra, Geller \& Corwin 1995; Geller \& Huchra 1989;
Huchra et al. 1990; Huchra et al. 1983; Davis et al. 1982) and SSRS
surveys (da Costa et al. 1994a; 1988) cover more than a third of the
sky and reach to a limiting apparent magnitude $m_B \sim 15.5$.

There is a rich literature analyzing these surveys to extract, for
example, the galaxy luminosity function (Marzke \& da Costa 1997;
Marzke, Huchra \& Geller 1994), the power spectrum for the galaxy
distribution (da Costa et al. 1994b; Marzke et al. 1995; Park et
al. 1994) and the velocity moments (Marzke et al. 1994). One factor
limiting these analyses is the inhomogeneous nature of the
databases. Here we take a step toward remedying that situation for the
CfA surveys which are based on the ZC (Zwicky \etal\ 1961-1968).

Measurement of redshifts for the CfA surveys began in 1978. The ZC
provided an obvious and, at that time, seemingly adequate source of
positions for galaxies in the northern hemisphere. The accuracy of the
Zwicky catalog coordinates ($3\arcmin$ at the
$3\sigma$ confidence level) was comparable with the
typical pointing accuracy of the Tillinghast Reflector on Mt. Hopkins,
the workhorse for the CfA surveys.  Nowadays, more comprehensive
scientific goals and significantly improved telescope pointing make
arcsecond coordinates imperative. Fortunately, the general
availability of the digitized POSS plates (DSS; Lasker et al. 1990)
enables us to revise the ZC by providing $\sim 2\arcsec$ coordinates
for each galaxy.  Here we provide these coordinates for \Nuzc\ 
galaxies with $m_{Zw} \leq 15.5$ that we were able to identify
unambiguously.

The measurement of a redshift for a nearby galaxy is now a rapid,
routine process. However, at the start of the CfA surveys, the much
slower pace made it seem judicious to accumulate redshifts from the
literature.  Given the effort required to construct the UZC, it is now
clear that it would have been better to remeasure the redshifts to
provide a uniform database. Approximately 30\% of the redshifts in the
catalog we describe here are from the literature; the rest were
measured at the CfA. Most of the CfA measurements were made with two
very different spectrographs, the Z-machine (Latham 1982) and FAST
(Fabricant et al. 1998).  We have re-reduced all of the Z-machine and
FAST data and have determined a uniformly calibrated error
(Kurtz \& Mink 1998; hereafter KM98).

The UZC contained here includes redshifts for 96\% of the galaxies
with m$_{Zw} \leq 15.5$ along with $\sim 1,400$ additional redshifts for
galaxies within multiplets in the region $20^h \geq \alpha_{1950} \leq
4^h$ and $8^h \geq \alpha_{1950} \leq 17^h$, both for $-2\pdeg5 \leq
\delta_{1950} \leq 50^\circ$ (where our completeness is 98\%;
hereafter we refer to this region as the CfA2 region).  We define UZC
multiplets as galaxies with magnitude differences $\Delta m < 0.5$ mag
and positional differences $\Delta \theta <3\arcmin$.  The catalog is
suitable for a wide range of statistical analyses.

Section 2 describes the contents of the original Zwicky-Nilson merged
catalog (ZNCAT). Section 3 describes the ZCAT compilation. Section 4
outlines our procedure for matching galaxies in ZNCAT and ZCAT with
the HST Guide Star catalog (GSC), as a first pass to obtain accurate
coordinates, and a subsequent refinement that improved the yield of
matched galaxies. Section 5 describes the CfA redshift observations
and reduction procedures. Section 6 outlines the procedures we used
for blunder identification and for blunder rate evaluation. Section 7
contains the catalog and comments on its possible applications, and
limitations.
 
\section{The Zwicky Catalog}

Our starting point for constructing the UZC is the database of
properties of galaxies in the ZC, the Zwicky-Nilson catalog (ZNCAT;
Tonry \& Davis 1979; hereafter TD79). ZNCAT was created in preparation
for the first CfA Redshift Survey (Davis et al. 1982); it covers the
entire northern sky and contains the union of 
CGCG galaxies in the Zwicky catalogs 
(Zwicky \etal\ 1961-1968) and 
UGC galaxies in the Nilson catalog (Nilson 1973). ZNCAT
contains position, magnitude and some morphology entries for 30,813
galaxies.  We restrict ourselves to the \Nzwi\ objects with m$_{Zw}
\leq 15.5$, the limit where Zwicky estimated that his catalog was
complete.  The CfA redshift surveys include redshifts for nearly all
of the galaxies to this limit over large areas.

Each galaxy in ZNCAT is described by a ``Zwicky Number'', another
catalog label such as NGC or IC, its B1950 coordinates, Zwicky's
estimate of its Johnson $m_B$ magnitude, m$_{Zw}$, and, if known, its
morphological type and its position angle on the sky and $B-$ and
$R-$band diameters in arcsec.  The errors in the galaxy magnitudes are
0.3 mag (Bothun \& Cornell 1990; Huchra 1976); the $3 \sigma$ error
circle for the galaxies has a radius of $\sim 3\arcmin$ (Zwicky et al.
1968).

\section{ZCAT}

ZCAT (Huchra et al. 1992) is a compilation of redshift data
for galaxies that currently contains approximately 100,000
entries. ZCAT includes positions, redshifts, magnitudes, and types
from a variety of sources.  We used ZCAT as an initial list of
redshifts to construct a complete catalog for part of the northern
sky.  The coordinates of $\sim 9,600$ objects in ZCAT are identical to
the corresponding ones in ZNCAT; thus, ZCAT coordinates are often
accurate to only $\sim 3\arcmin$ ($3\sigma$).  Revised coordinates in ZCAT
frequently remain good to $\sim 1\arcmin$ ($3\sigma$) at best. There is also
possible confusion with neighboring bright galaxies in $\sim 15\%$ of
the entries. Such confusion cannot be eliminated without both accurate
coordinates and magnitudes (see de Vaucouleurs et al. 1991; hereafter RC3). 
Without the latter, the only alternative
was to determine accurate coordinates as we do here, and remeasure
redshifts based on these coordinates. We recently began a program of
redshift remeasurement; we include the current results of this program
in the UZC.

Magnitudes from the ZC in ZCAT are problematic.  In the ZCAT
compilation, the magnitudes from the original ZC are often replaced
with measurements from a multitude of other sources.  In cases of
``multiplets'' defined as for the UZC, ZCAT often contains eye
estimates of split magnitudes. For the UZC, we restore the original
Zwicky magnitudes $m_{Zw}$ because they are the only system uniformly
available for all of the galaxies and because they determined the
original selection of galaxies for inclusion in the CfA surveys. In
the catalog, we flag these multiples; the split magnitudes are
available for reference in ZCAT (Huchra \etal\ 1992).  The arcsecond
coordinates we provide make it easy to measure and/or include better
magnitudes if more reliable sources of photometry are available
(see also RC3).

In using the UZC for statistical analyses, it should be noted that the
flagged magnitudes of multiplets could be systematically too bright for
each galaxy at the coordinates listed in the catalog.  A careful
reading of Zwicky et al. (1961) reveals that magnitudes in the Zwicky
catalog were not compiled for multiplets in a uniform fashion.  In
cases where the multiplet is unresolved, the magnitude listed by
Zwicky et al. (1961) is probably too bright for any single
component. In resolved cases, the listed magnitude may be that for the
brightest member alone. In the absence of a clear indication of the
procedure of Zwicky et al. (1961), the safer alternative 
that we followed was to list only the original Zwicky et al. (1961)
magnitude along with 
the brightest member.  Furthermore, ours is not an
exhaustive catalog of multiplets: those listed in the UZC were included
because the magnitudes ($m_{Zw}$) of the components differed by less
than $\sim 0.5$ mag. However, the UZC is a source for further
searches for and statistical studies of multiplets.

\section{Matching Catalogs to Obtain Arcsecond Positions}

We compared the positions of galaxies in ZNCAT and ZCAT with an
independent source of accurate coordinates, the HST Guide Star Catalog
(GSC). The GSC contains coordinates for stars, and for extended
objects (``non-stars'').  Typical accuracies for all GSC coordinates
are $\sim 1\arcsec$.  However, the main focus of the GSC was on $V <
14$ mag stars. Thus, the magnitudes and identifications of galaxies
are less reliable than for the stars (e.g., Lasker et al. 1990;  
Alonso et al. 1993, 1994). With
these limitations in mind, we selected objects from the GSC which
match the positions of the objects in ZNCAT.

We matched GSC and ZNCAT positions with software written for this
task.  From a list of ZNCAT and GSC coordinates, we calculated the
Cartesian distance on the sky between objects in ZNCAT and in the GSC,
and searched for matches within a $6\arcmin\times6\arcmin$ box
centered on each ZNCAT position; we found 19,878 matches.  Out of
these, 1,854 were duplicates due to overlaps of the digitized sky
survey plates, for a total of 18,024 non-duplicate matches.

We made $6\arcmin\times 6\arcmin$ finding charts extracted from the
DSS, centered on each original ZNCAT position. Our software classified
these charts as follows:

\begin{enumerate}

\item class Z: a GSC class 3 (a ``non-star'') object appears within a
$6\arcmin\times6\arcmin$ box centered on the Zwicky position.  For the
16,268 fields in this category, there is a high likelihood that each
GSC object is the Zwicky galaxy.

\item class N: no GSC class 3 object appears within
$6\arcmin\times6\arcmin$ box centered on the Zwicky position, but one
GSC class 0 object (a ``star'') is within the box.  For the 3,363
fields in this category, the object nearest the Zwicky position is
probably a star in the GSC.

\item class B: no GSC object of any class appears within the
$6\arcmin\times6\arcmin$ box centered on the Zwicky position.  There
are 247 fields in this category. Here, either through an error in the
GSC classification, or through an error in the ZNCAT coordinates,
there is no obvious candidate galaxy at the ZNCAT position.

\end{enumerate}

We printed DSS finding charts for all the objects in our
sample. Charts of type Z and N are centered on the matching GSC
coordinates.  Charts of type B are centered on the ZNCAT coordinates.
To each matched object, we assigned a running index increasing with
apparent magnitude. Ordered in this way, objects of similar apparent
brightness are sequentially indexed close together. Thus, the visual
classification described below remains consistent as a function of
decreasing apparent brightness.

We also matched objects in ZCAT and ZNCAT by searching each
$6\arcmin\times6\arcmin$ coordinate box (centered on each ZNCAT object
with $m_{Zw}\leq 15.5$ mag) with corresponding objects in ZCAT (with no
restriction on the magnitudes). We found 19,584 matches, a number 
exceeding our total count (18,024) of matched non-duplicate ZNCAT
objects within our magnitude limit. The mismatch occurred because (1)
there were multiple matches within our matching box 
and (2) there were matches with ZCAT
objects fainter than our choice of magnitude limit.

Because there are significant errors in both the GSC classification
and in the ZNCAT coordinates, we could not be certain that
our Z or N-class field classifications corresponded to the actual
Zwicky galaxies. However, we had a preliminary ranking of galaxies in
order of decreasing likelihood (from Z to N to B) of matching a ZNCAT
galaxy.  We therefore examined all of the Z, N and B fields visually,
to determine a second classification of the objects. This
time-consuming classification substantially improved the likelihood
that a matched object in the GSC is actually a ZNCAT galaxy.  The new
classification consisted of numerical indices 0$-$4:

\begin{enumerate}

\item{} code 0: a ``hit'', where the GSC and ZNCAT objects are
separated in position by $\Delta\theta\leq 180\arcsec$, are comparable
in brightness as determined visually, and there is no other object
within the field that could be a ZC galaxy. Our indexing scheme
furnished a qualitative estimate for the range of magnitudes expected
for each galaxy, which we used as a guide during the visual
inspection.  Note that the magnitudes m$_{Zw}$ (B band) and m$_{\rm
GSC}$ (V or J bands) can differ significantly, because the expected
colors for these galaxies are in the range $\sim 0.3-1.0$ mag
(Frei \& Gunn 1994).
 
\item{} code 1: a ``hit'' with $\Delta\theta\leq 180\arcsec$, again with a
GSC brightness in the vicinity of the ZNCAT value. However, the field
is centered on a GSC ``star'' rather than on a ``non-star'', but there
is only one possible ZC galaxy nearby, within the field. Thus, index 1
merely indicated that the nearest GSC object to the ZCAT coordinates
is a star rather than the appropriate galaxy;

\item{} code 2: a near ``hit'' with $\Delta\theta\leq 180\arcsec$.  However,
there is confusion because there are at least 2 galaxies with an
estimated spread in magnitudes $|\Delta m|$ within
the expected range, either of which could be a ZC galaxy;

\item{} code 3: also a near ``hit'' with $\Delta\theta\leq 180\arcsec$
but with $|\Delta m|$ outside the expected range, raising suspicions
about the match of the GSC object to the ZC galaxy;

\item{} code 4: no ``hit'' at all for $\Delta\theta\leq 180\arcsec$, i.e.,
there is no match for a ZC galaxy.

\end{enumerate}

One of us (B. Elwell) examined each chart and assigned it a code of
$0-4$, according to these criteria.  Our strategy was to print finding
charts in order of increasing magnitude; thus, we always knew the
range of brightness being examined. 

The visual examination yielded 
refined, accurate estimates for the appropriate object to
match to the ZC coordinates. There were 3 distinct outcomes:

\par\noindent 
1) For perfect ``hits'', there is no doubt that the Zwicky galaxy is
matched.  The coordinates are those of the matching object in the
GSC. We examined our printed DSS finding charts to confirm the match
visually.  These $12,154$ galaxies are in class 0 above.

\par\noindent
2) For near ``hits'' in class 1, there is no doubt that we have a match for
the ZC galaxy, but the GSC coordinates did not fall within $\sim
2\arcsec$ of the center of brightness of the ZC galaxy.  One of us
(J. Peters) confirmed the identification with software which allowed
interactive examination of the DSS finding charts for the coordinates
of each ZC galaxy. These $2,799$ galaxies were in class 1 above.

\par\noindent 
3) For confused ``hits'' in classes $2-4$, there is more than one
possible match.  In these cases, we have 3 sources of confusion: (a)
at least 2 objects of approximately the expected magnitude lie within
the 3\arcmin\ error circle; (b) the nearest match to the ZNCAT
position has the wrong magnitude (invariably fainter); and (c) the
3\arcmin\ error circle is empty.  We found $4,631$ galaxies in these
classes, leading to a second round of visual examination
using software in the WCStools package (Mink 1999). We examined
the original finding charts used for the redshift measurements to
associate the galaxy with its ZC position and with its measured
redshift.  When a match was found, the coordinates were determined
interactively, with WCStools software.
The success rate for this procedure was low, only $\sim 30$
\%, because of the non-uniformity of the record-keeping, and of poorly
marked or missing charts. Furthermore, we found many cases ($\sim
20$\% of the charts we examined) where the redshift in ZCAT did not
match the entry on the finding chart.

The total number of galaxies in classes 0-4 matches the total count of
19,584. At this stage we eliminated duplicates and thus pared the
total number of entries to the number of unique ZNCAT galaxies. This
total number is the count of galaxies in ZNCAT with non-zero
magnitudes after eliminating (1) a small number of duplicate entries;
(2) 19 entries for which no clear galaxy was found within $15\arcmin$ of the
Zwicky position; and (3) the matches for 063024$+$44480 (NGC2242), a known
planetary nebula, and for 065012$+$16590, a known open cluster, both of 
which were mistakenly included in the ZC. The completeness
values reported above are slightly higher if we take into account the 
ZC entries that we eliminated.

\section{Redshifts in the UZC}

The redshifts and errors listed for CfA data here take precedence over
values published previously, including those measured with the
Z-machine.  All the CfA redshifts in the UZC were determined from
observations made at Mt.  Hopkins with the MMT, or with the Z-machine
or FAST spectrographs on the 1.5-meter Tillinghast reflector at the
Whipple Observatory.

We include in the UZC 4,465 redshifts measured with FAST. Of these,
1,905 are previously unpublished redshifts in the CfA2 region, and the
rest are in the remaining regions of the northern sky, as well as
remeasurements. About 1,000 of the remeasurements were necessary in
cases where there was ambiguity in the identification of unresolved
galaxies even after looking at historical records.  Our source of UZC
redshifts from the literature was ZCAT; about 25\% of the redshifts in
ZCAT were compiled from the literature. In the CfA2 survey region, CfA
measurements make a slightly larger contribution, with 1.4\% of the
redshifts from the MMT, 61\% from the Z-machine and 17\% from FAST.

The Z-machine (Latham 1982) measurements were made between 1978 and
1993.  For these observations, a 600 line mm$^{-1}$ grating yielded a
resolution of 5\AA\ in first order over the wavelength range 4500 -
7100 \AA. The typical integration time for each object was 15-50
minutes.  In 1993, the FAST spectrograph (Fabricant et al. 1998) was
first mounted on the 1.5-meter Tillinghast. The FAST spectra have 6
\AA\ resolution and coverage of 3700 - 7500 \AA. Typical integration
times were 3-10 minutes.

For our FAST observations, when there was more than one galaxy near
the ZNCAT position and within $\sim 0.5$ mag of the original Zwicky
magnitude, we measured redshifts for all of the galaxies. We began the
observing program in September 1996, and completed it in October
1997. All of these new redshift measurements are included in the UZC
(see Table 1). We also updated the coordinates of all of these
remeasured galaxies, using the DSS finding charts, with an estimated
positional uncertainty of $\sim 2\arcsec$.  FAST measurements (from
our own and from other projects) replaced any previously measured
redshifts, e.g. with the Z-machine, in the UZC.

\section {Reliability of the Redshifts in the UZC}

In many cases where the GSC galaxy is an unambiguous match to ZNCAT,
we checked our DSS finding charts against those used for the CfA
Redshift Survey.  We corrected obvious typos found in this way. This
procedure underscored our awareness of inadequacies in the reduction
and record-keeping, and spurred us to make an objective {\it
measurement} of the error rate in redshifts from the literature and
from our own facilities by remeasuring redshifts for a significant
number of objects chosen with a variety of criteria.

We first used our current archive of repeat measurements with FAST to
estimate the error rate in these new data. For 620 pairs of FAST
measurements, there were no blunders (KM98).  Thus, we only have an
upper limit of $\sim 0.2\%$ for the error rate, which is attributable
to the much better pointing of the Tillinghast 1.5-meter, much
improved observing protocols, and better record-keeping and archiving
procedures. We are thus confident that for the 3,446 FAST redshifts in
the UZC, the blunder rate is essentially zero.

We reduced the 4,366 FAST spectra of galaxies in our sample in a
uniform manner, using the IRAF task {\bf rvsao} (KM98).  All spectra
were cross-correlated with two templates, fabtemp97, a composite
absorption line spectrum (KM98), which is generally used for all FAST
redshifts, and femtemp97, a synthetic emission line template (KM98).
We estimated the errors for our FAST spectra as in KM98: the values
are obtained by dividing a constant by $1+R$ ($R$ values are a measure
of the quality of fit of a template and of S/N for the spectra, see
TD79, KM98), and adding $20/\sqrt{2}$ \kms\ in quadrature.  For
fabtemp97 the constant is 350 $\kms$, and for femtemp97 the constant
is 220 $\kms$.  These constants were selected from Figures 10 and 12
of KM98, to optimize the contributions of statistical and systematic
errors from cross-correlation with each template.  We required $R>3$
for all the redshifts we report here. In cases where the difference
for the two templates is less than 300 \kms, the redshift in Table 1
is the error-weighted combination of the two, as is its error. Of the
4,366 spectra, 30 were of insufficient quality, 1,872 matched the
absorption line template fabtemp97, 692 matched the emission line
template femtemp97, and 1,772 matched both templates.

We also re-reduced the 10,051 Z-machine spectra of the galaxies
in our sample in a uniform manner with {\bf rvsao} (KM98).  All spectra
were cross-correlated with the templates ztemp, a composite
absorption line spectrum that has been used for all Z-machine
redshifts since TD79, and femtemp97. 
Figure 1 shows spectrograms of ztemp and femtemp97 (compare
with Figure 11 of KM98).

The Z-machine spectra do not have sufficient quality to permit fully
automatic reduction; the initial reduction required visual inspection
of every spectrum.  We compared the new redshifts, obtained by
correlation with the absorption and emission line templates, with
those from the original reduction; we also checked for sufficient
signal to noise, requiring $R>3$.  We accepted a redshift if it fell
within 300 $\kms$ of the original reduction; of the 10,080 Z-machine
spectra we accepted 9,936. We found that 4,424 matched with the
absorption line template, ztemp, 2,507 matched with the emission line
template femtemp97, and 3,005 matched both templates.  Among the
remaining 156 low-S/N spectra (FAST and Z-machine), 125 are in our 98\%
complete region. These galaxies are flagged in Table 1 (see below); we
plan to update their redshifts with FAST.

In the cases where a single template matches the original reduction,
the redshift in Table 1 is that obtained from that template, and the
error is calculated with the methods outlined in KM98. The Table also
indicates which template matched best, with a label of A for ztemp, E
for femtemp97, and B for both.  We calculated the error as for FAST
but with appropriate constants, again by dividing a constant by $1+R$,
and adding 35 $\kms$ in quadrature.  For ztemp the constant is 212
$\kms$, and for femtemp97 the constant is 140 $\kms$; these constants
are $3/8$ of the median width of correlation peaks where $R>4$ (KM98).
In cases where both templates match the original reduction, the
redshift in Table 1 is the error-weighted combination of the two, as
is the error.  In cases where the match with the original reduction
failed, we estimated errors individually to match the system defined
by successful cross-correlation.  ZCAT contains another 670 low-S/N
spectra that would be included in the UZC if their S/N were
sufficiently high. Any galaxy that was identified unambiguously but
whose redshift remains unknown is entered in Table 1 with accurate
coordinates, with nothing in its redshift column. Galaxies for which
spectra were obtained at CfA but where the S/N was below our
acceptance criterion are flagged in Table 1 (see below).

Even after the new measurements for ambiguous cases, the blunder rate
in ZCAT is significant for the Z-machine data and for data from the
literature.  For evaluation of the reliability of redshift
measurements including those obtained with the Z-machine and those
compiled from the literature, we extracted a random sample of 129
galaxies from ZCAT and re-measured their
redshifts with the FAST spectrograph.  We also assembled several
samples of FAST redshifts acquired for other projects: the Century
Survey (106 galaxies: Geller et al. 1997); studies of groups of
galaxies (66 galaxies: Mahdavi 1998, private communication); galaxies
in voids (226 galaxies: Grogin et al.  1998); and other galaxy surveys
(315 galaxies: Carter 1998, private communication).  The total number
of re-measured redshifts is 843.  In the UZC, we have replaced the
ZCAT redshift with the corresponding FAST measurement in all cases
(the number of FAST redshifts discussed above includes these
measurements and re-measurements).

For each sample, Figure 2 shows the differences between the FAST
measurements of redshifts and the corresponding original ZCAT entry,
plotted against the FAST redshifts represented as $cz$.  The
``blunder'' rate for the overall test sample, as well as for the
individual samples, is stable at $\lesssim 3$ \% for measurements not
made with FAST.  We find the same rate regardless of the origin of the
redshift in ZCAT (the literature, or Z-machine for example).  An oddly
large number of redshifts in ZCAT from the literature differ by 1,000
\kms\ from FAST measurements (only one of which happened to be included
in Figure 2), making us suspect typographical errors or confusion of
H$\alpha$ (6563\AA) with [NII] (6548\AA, 6584\AA) emission lines as 
major sources for the discrepancy.


\section{Discussion}

We have assembled a new version of the Zwicky Catalog (ZC), the
Updated Zwicky Catalog (UZC), with a magnitude limit of $m_{Zw}= 15.5$
mag. The UZC is a 98\% complete redshift catalog to this magnitude
limit, in the CfA2 region with $20^h \geq \alpha_{1950} \leq 4^h$ and
$8^h \geq \alpha_{1950} \leq 17^h$ and $-2\pdeg5 \leq \delta_{1950}
\leq 50^\circ$.  The completeness of the UZC is lower at the northernmost
declinations, e.g., if we restrict $\delta_{1950}\ge 50\deg$, it
falls to 90\%.  The main advantages of the UZC over previous
compilations are (1) uniformly accurate coordinates at the $\leq
2\arcsec$ level; (2) a robust estimate of the accuracy of the CfA
redshifts in the range $cz = 0-25,000\kms$, with a current total of
$\sim 25$\% of redshifts with essentially no ``blunders''; (3) an
estimate of the reliability of catalog redshifts for data from the
literature and from the Z-machine: the remaining 75\% of the redshifts
have a blunder rate of $\ltorder 3$\%.  A continuing problem with the
UZC is that magnitudes for the galaxies in the sample are still the
original Zwicky mgnitudes, which have $\sim 0.3$ mag errors. Multiples
are flagged in the UZC (see Table 1), because we have not split the
Zwicky magnitudes of their components. These magnitudes may be 
systematically too bright for each component of an 
unresolved UZC multiplet (\S 3).

Table 1 contains a sample listing of the UZC.  Column 1 (``RA (J2000)
DEC'') shows the right ascension and declination (J2000) of each
galaxy. Column 2 (``{\it B}'') shows the $B$-band magnitude $m_{Zw}$
from ZNCAT. Columns 3 (``c$z$'') and 4 (``$\Delta$c$z$'') show the
heliocentric 
redshift and its error in \kms.  Column 5 (``T'') indicates the type
of the redshift for each galaxy, E (emission), A (absorption) or B
(both emission and absorption).  Column 6 (``U'') shows the the UZC
code (0-4) assigned to each galaxy, according to the quality of the
match with DSS galaxies (see \S 4).  Column 7 (``$N$'') shows the
number of UZC neighboring galaxies to the current entry, within
3\arcmin.  Galaxies in the UZC with a number of neighbors larger than
0 are thus in UZC ``multiplet'' systems. All entries with $N>0$ are
flagged with a star in Column 12, to indicate membership in a UZC
``multiplet''.  Column 8 (``ZNCAT Name'') shows the ZNCAT or Zwicky
label; multiplets generally share a single such label.  Column 9
(``Ref. code'') indicates the origin of the redshift measurement, for
data taken at the CfA: Z for Z-machine, M for the MMT and F for FAST
spectra taken for our (other) project(s).  The label ``X'' flags a
low-S/N spectrum for Z-machine (FAST) measurements and
indicates a low S/N match with a spectrum at c$z<100$ \kms,
an indication that no redshift could be determined. 
Column 10 (``Ref.'')  shows a ZCAT literature reference number (in
turn listed in Table 2), or a blank for CfA unpublished data.  Column
11 (``Other Name'') shows any other name may be associated with the
current entry, such as an NGC number, as listed in ZCAT.  Column 13
indicates whether the galaxy is a member of a Zwicky multiplet, as
reported by NED (``P'', ``T'' and ``G'' for pairs, triples and groups,
respectively).  Because of the size and limited utility of a printed
version, the full table is available only with the electronic version
of the journal article.  
This table, as well as additional measurements we will carry out
with FAST will also be available at URL http://tdc-www.harvard.edu.

Figure 3 shows the distribution of the magnitude of the differences
between UZC and Zwicky coordinates. The mode of the histogram is at
$\sim 40\arcsec$, slightly better than the $1\arcmin$ ($1\sigma$)
errors claimed by Zwicky. For additional confirmation of our
positional accuracy, we matched the FIRST 1.4 GHz catalog (White et
al. 1997) with the UZC, using a search radius of 10\arcsec.  We found
matches for 1,347 FIRST sources. The distribution of coordinate
differences for these sources has a narrow peak at $\sim 1\arcsec$,
with a width $\sigma = 1\farcs45$. Thus, for these FIRST matches, the
UZC positional uncertainty is $< 1\farcs5$, which confirms the
significant improvement in positional accuracy that we have achieved
with the UZC.

Figure 4 shows the sky distribution of UZC galaxies 
with measured redshifts, as listed in Table 1. 
The UZC covers only the northern sky; regions devoid of galaxies
result from obscuration by the Milky Way.  Figure 5 shows the
sky distribution of galaxies without measured 
redshifts, also as listed in Table 1. A comparison of
Figures 4 and 5 shows that there are no significant patterns in the
distribution of galaxies without measured redshifts, other
than, e.g., those north of $+50\arcdeg$ and
south of the equator, with a known lack of coverage.  Figure 6 shows
cone diagrams for the distribution of galaxies in the UZC as a
function of $\alpha$, $\delta$ and c$z$ in \kms. Each ``slice''
spans $\sim 10\arcdeg$ of declination and right ascension
ranges of $8-17h$ in the northern galactic polar region and $20-4h$ in
the South galactic cap. These slices cover the CfA2 region of the
redshift survey.  

The known structure of the distribution of galaxies is readily
apparent in these diagrams (Geller \& Huchra 1989).  The large voids
and coherent sheet-like structures appear in adjacent sets of the
``slices''.  Figure 7 shows histograms of the number of galaxies
$N(z)$ as a function of redshift for both polar galactic regions in
the UZC.  The structure known as the Great Wall (Geller \& Huchra
1989) that persists at $7,000-10,000$ \kms\ in each of the $8-17h$
slices in Figure 6, as well as in Figure 7. The latter figure shows
significant peaks at the redshifts of the Great Wall and the Virgo
cluster in the North, and at the redshift of Perseus-Pisces in the
South galactic hemisphere. Such peaks confirm the presence of these
large-scale structures, and demonstrate their narrowness in redshift
space. These in turn are characteristic of the sheet-like structures
in Figure 6.

We thank E. Barton, B. Carter, N. Grogin and A. Mahdavi for use of
their unpublished data. We also thank D. Fabricant for comments and
J. Roll for comments and indispensable assistance with the
software. We also thank the referee, Harold Corwin, for useful
suggestions which improved the clarity of our presentation of the UZC.
This research was supported in part by the Smithsonian Institution.

\newpage
\clearpage
\figurenum{1}
\begin{figure}
\plotone{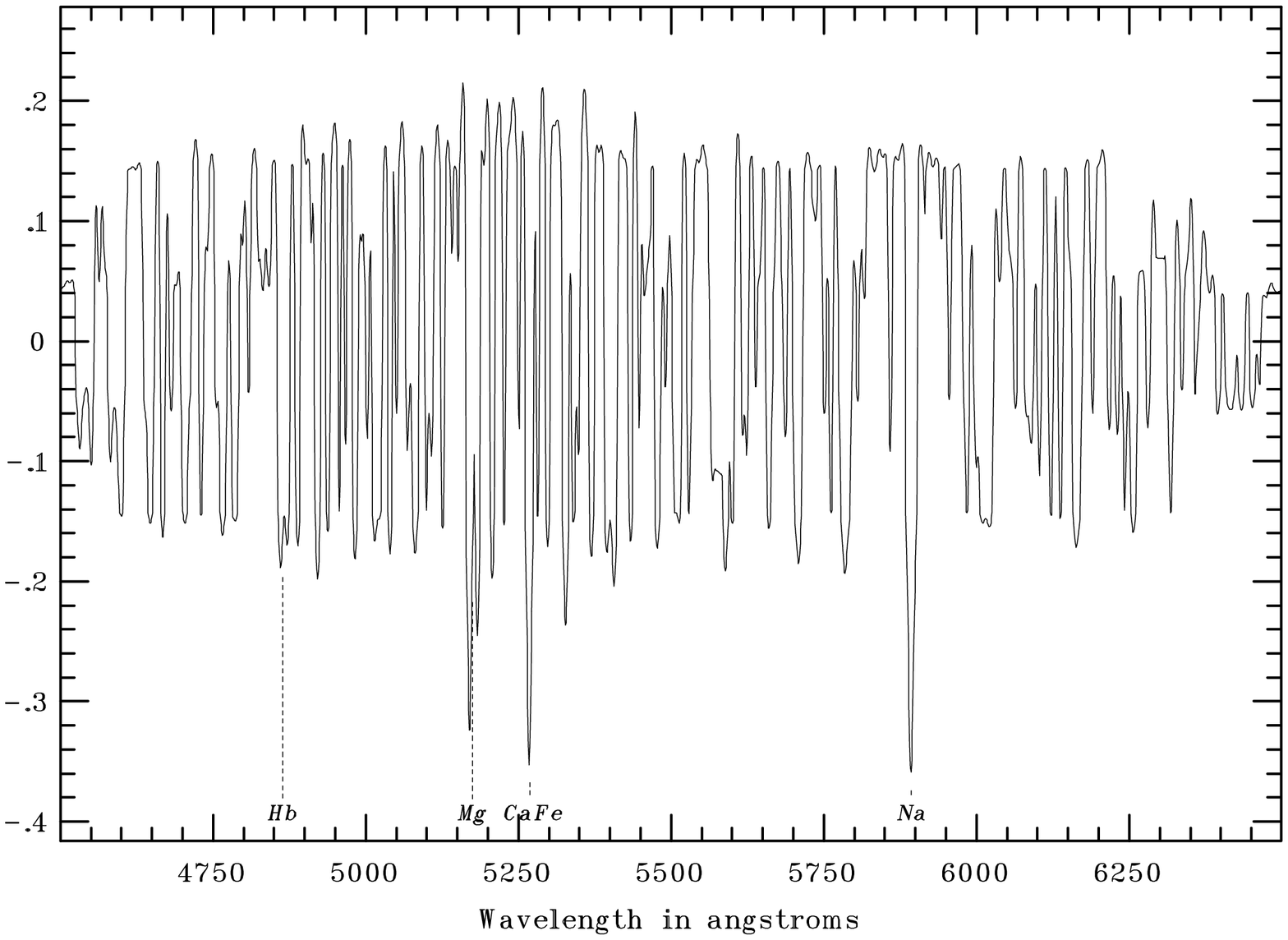}
\figcaption{
(a) The absorption template ztemp. 
Prominent metal and Balmer absorption lines are labeled. 
The less prominent but prevalent absorption lines in
the absorption template are real features.
}
\end{figure}

\newpage 
\clearpage
\figurenum{1}
\begin{figure}
\plotone{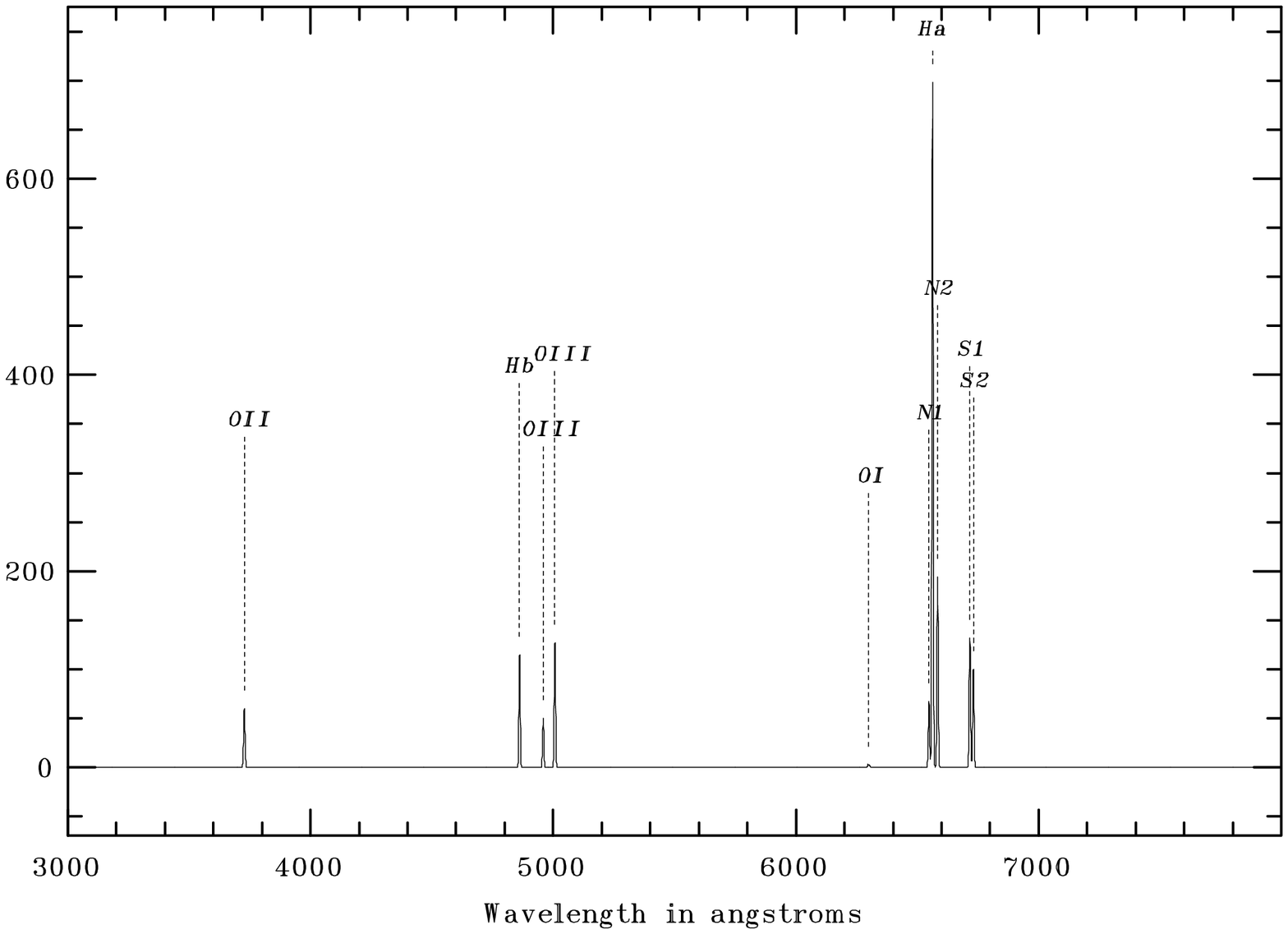}
\figcaption{
(b) The emission template femtemp97. Prominent metal and Balmer 
emission lines are labeled. 
}
\end{figure}
\newpage 
\clearpage

\figurenum{2}
\begin{figure}
\plotone{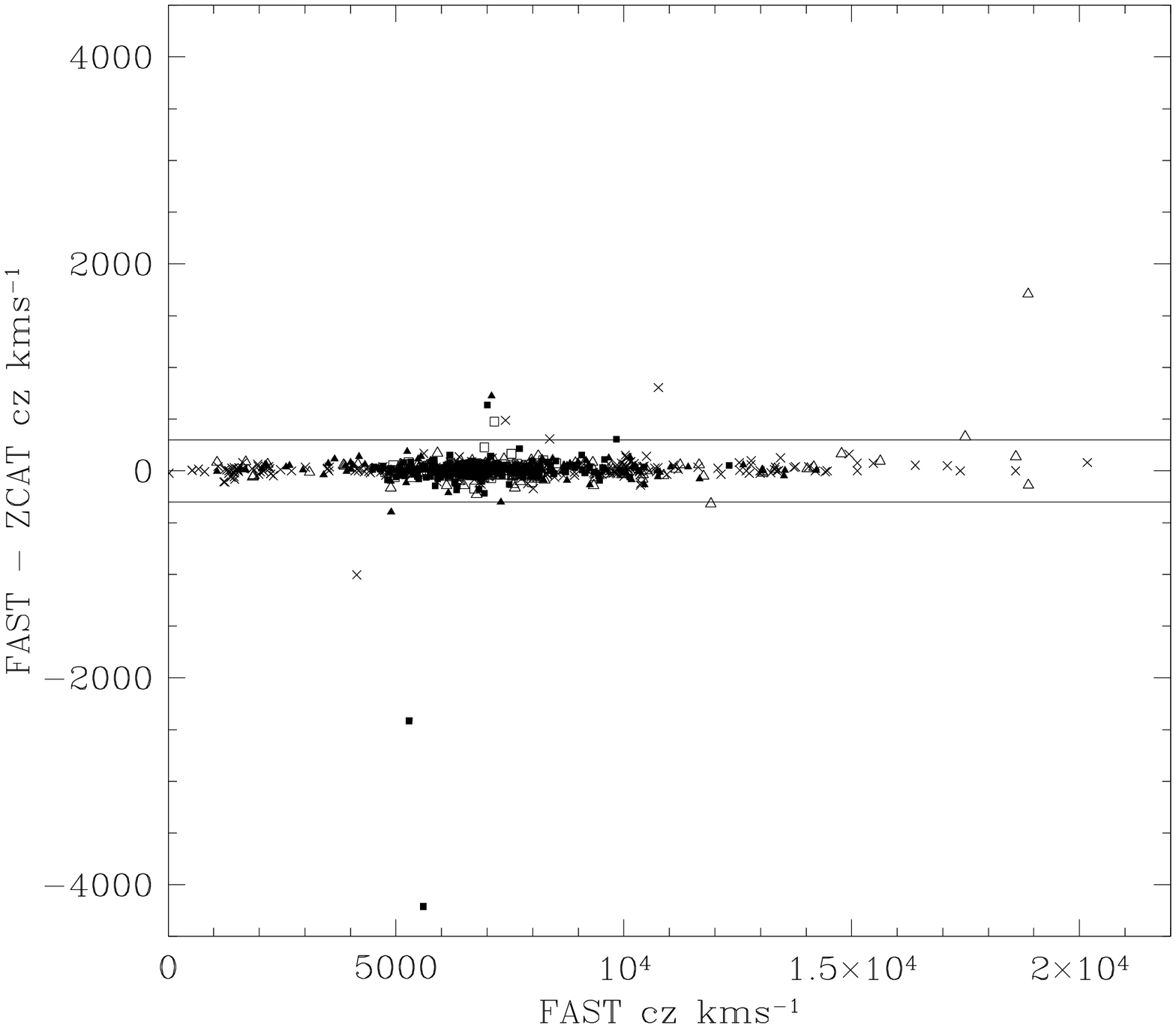}
\figcaption{
The differences between ZCAT and FAST redshifts, plotted as a function
of FAST redshift.  The symbols are: open triangles (Century Survey),
filled triangles (Comparison survey), open squares (Mahdavi survey),
filled squares (Grogin survey), and crosses (Carter survey).  We chose
the limits for clarity; therefore, a few ``blunders'' are outside the
plot. Horizontal lines at $\pm$ 300 \kms\ are intended to guide the
eye and provide a comparison across the plot.
}
\end{figure}

\newpage 
\clearpage
\figurenum{3}
\begin{figure}
\plotone{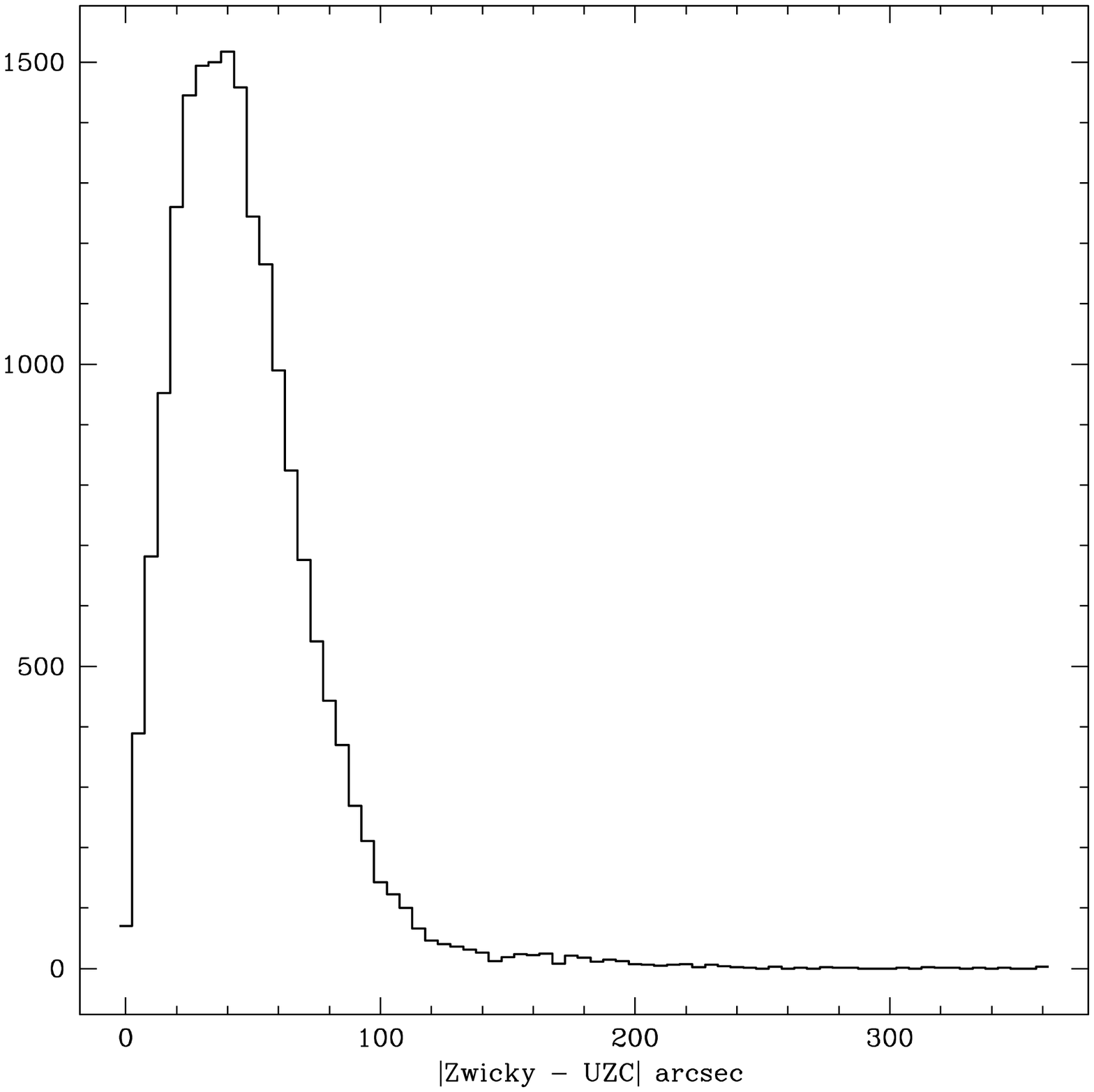} 
\figcaption{ 
The distribution of the absolute value of the
differences between UZC and Zwicky coordinates.
}
\end{figure}

\newpage 
\clearpage
\figurenum{4}
\begin{figure}
\plotone{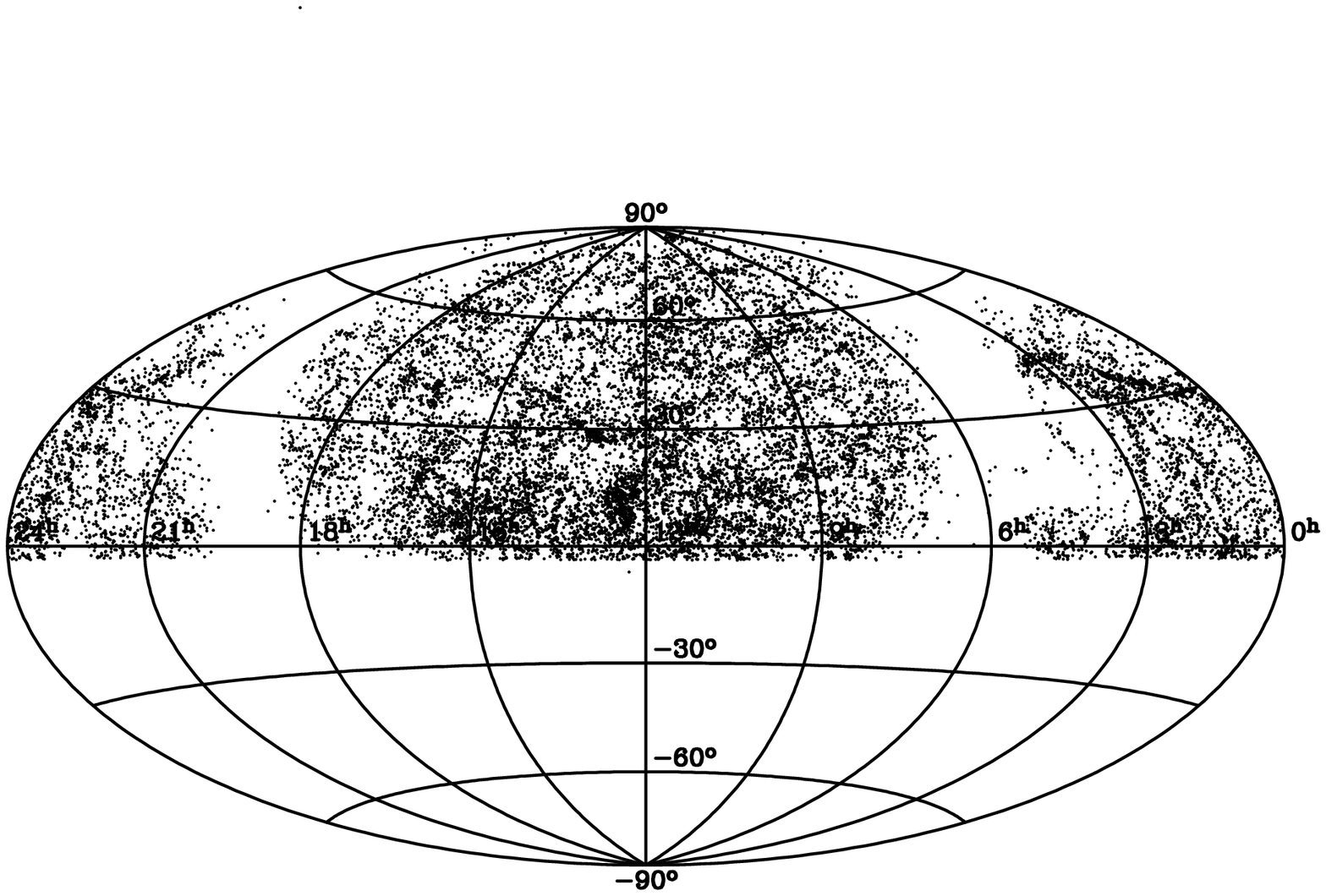}
\figcaption{
The distribution on the sky of the UZC galaxies. The Galactic plane 
appears as an empty band. 
}
\end{figure}

\newpage 
\clearpage
\figurenum{5}
\begin{figure}
\plotone{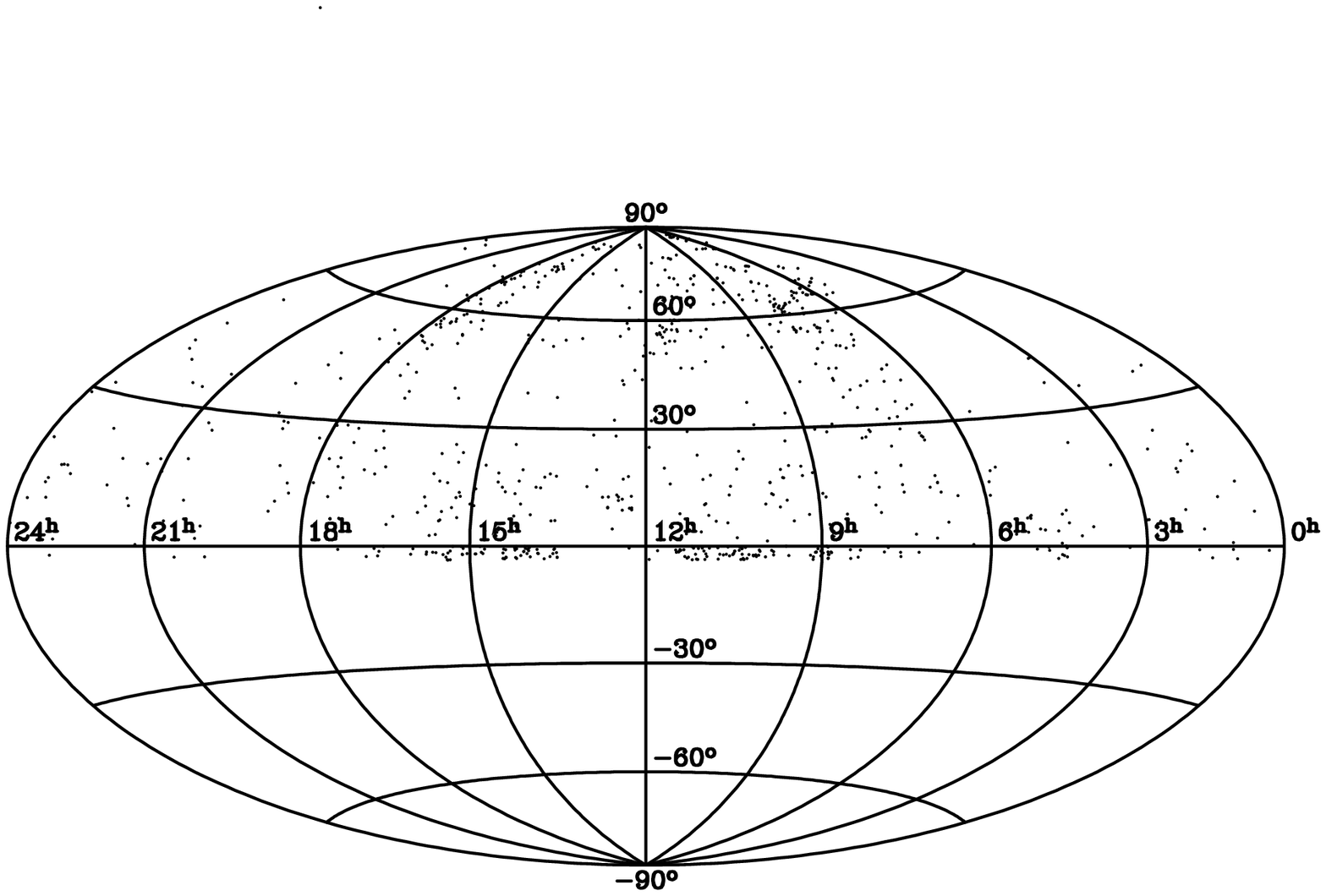} 
\figcaption{ The distribution on the sky of Zwicky
galaxies without measured redshifts in the UZC.  
}
\end{figure}

\newpage 
\clearpage
\figurenum{6}
\begin{figure}
\plotone{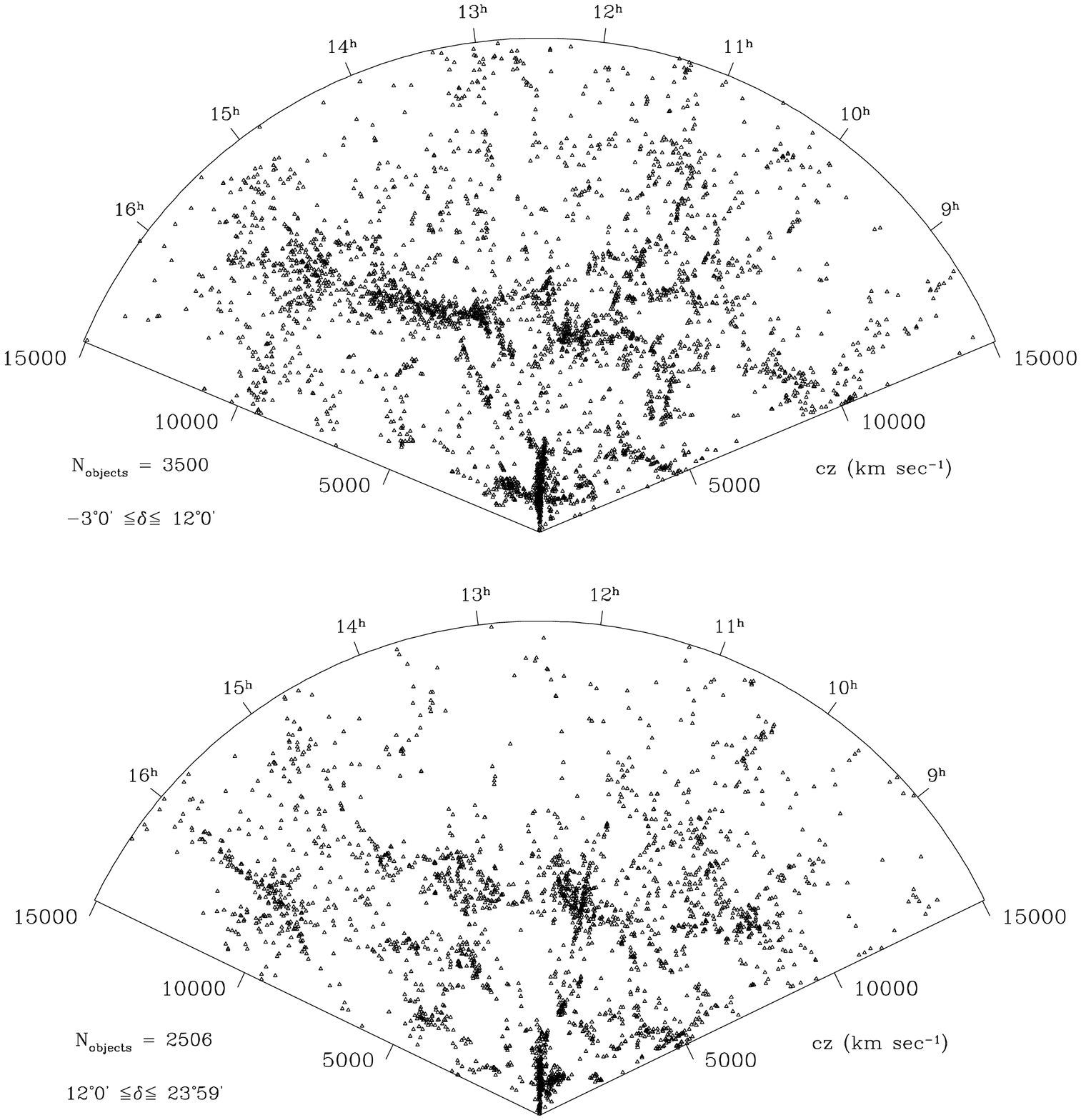}
\figcaption{
(a) Cone diagrams for $8 \leq \alpha_{1950}\leq 17h$. Each 
UZC galaxy is represented by a small triangle. The thickness of 
each ``slice'' is $\Delta\delta_{1950}\approx 12\deg$.
}
\end{figure}

\newpage 
\clearpage
\figurenum{6}
\begin{figure}
\plotone{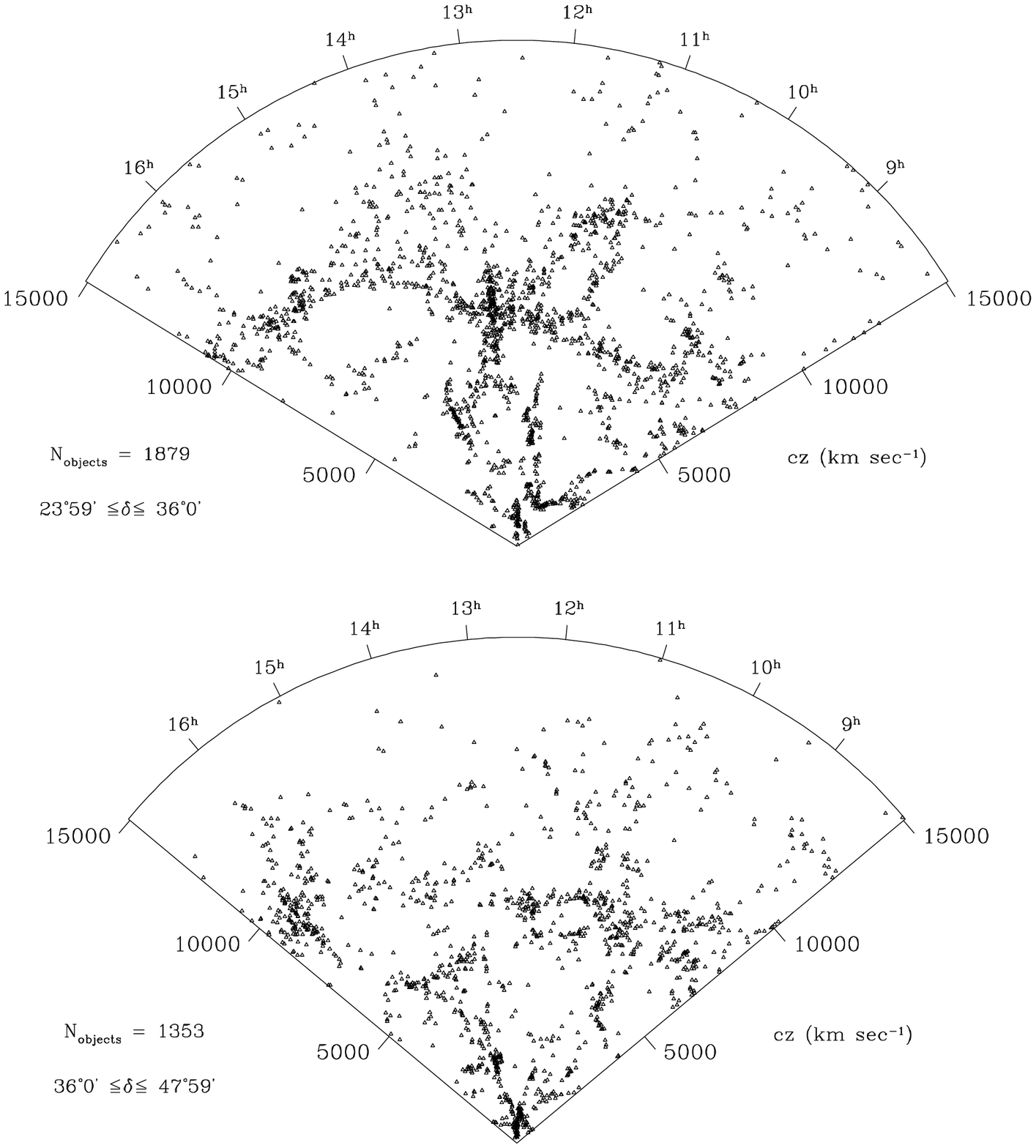}
\figcaption{
(b) Cone diagrams for $8 \leq \alpha_{1950}\leq 17h$. Each 
UZC galaxy is represented by a small triangle. The thickness of 
each ``slice'' is $\Delta\delta_{1950}\approx 12\deg$.
}
\end{figure}

\newpage 
\clearpage
\figurenum{6}
\begin{figure}
\plotone{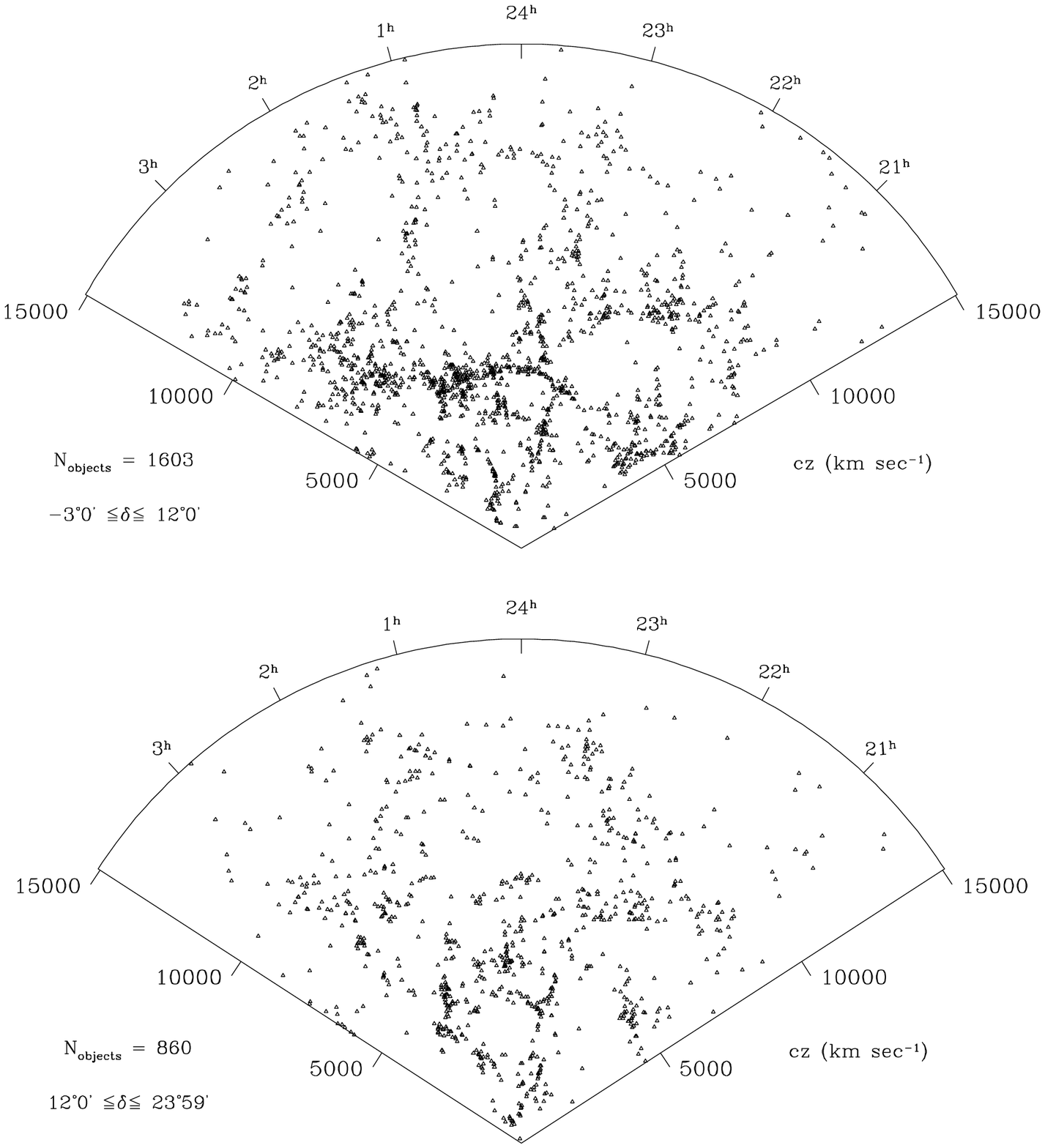}
\figcaption{
(c) Cone diagrams for $20 < \alpha\leq 24h$ and $0 < \alpha\leq 4h$. 
Each UZC galaxy is represented by a small triangle. The thickness of 
each ``slice'' is $\Delta\delta_{1950}\approx 12\deg$.
}
\end{figure}

\newpage 
\clearpage
\figurenum{6}
\begin{figure}
\plotone{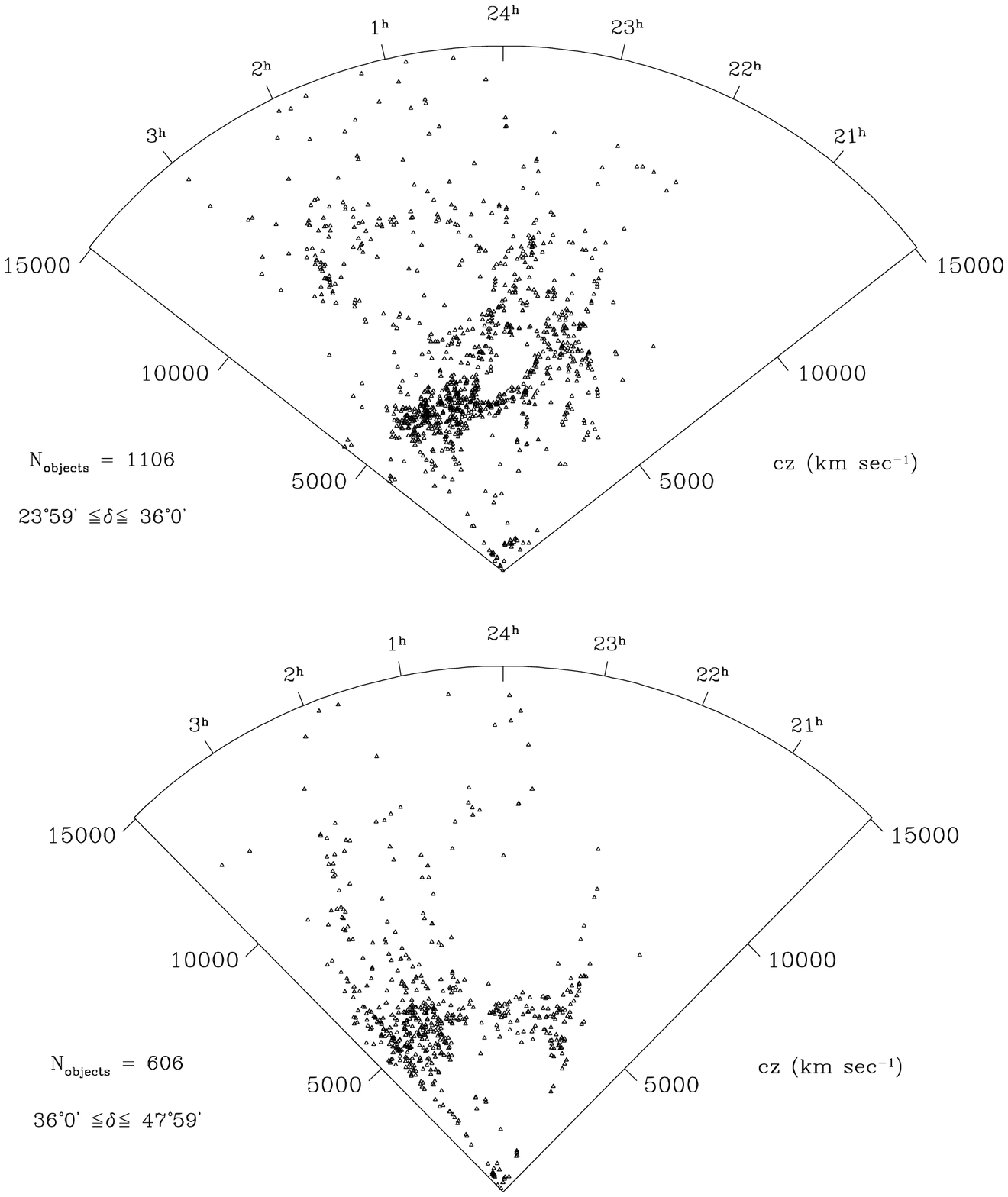}
\figcaption{
(d) Cone diagrams for $20 < \alpha\leq 24h$ and $0 < \alpha\leq 4h$. 
Each UZC galaxy is represented by a small triangle. The thickness of 
each ``slice'' is $\Delta\delta_{1950}\approx 12\deg$.
}
\end{figure}

\newpage 
\clearpage
\figurenum{7}
\begin{figure}
\plotone{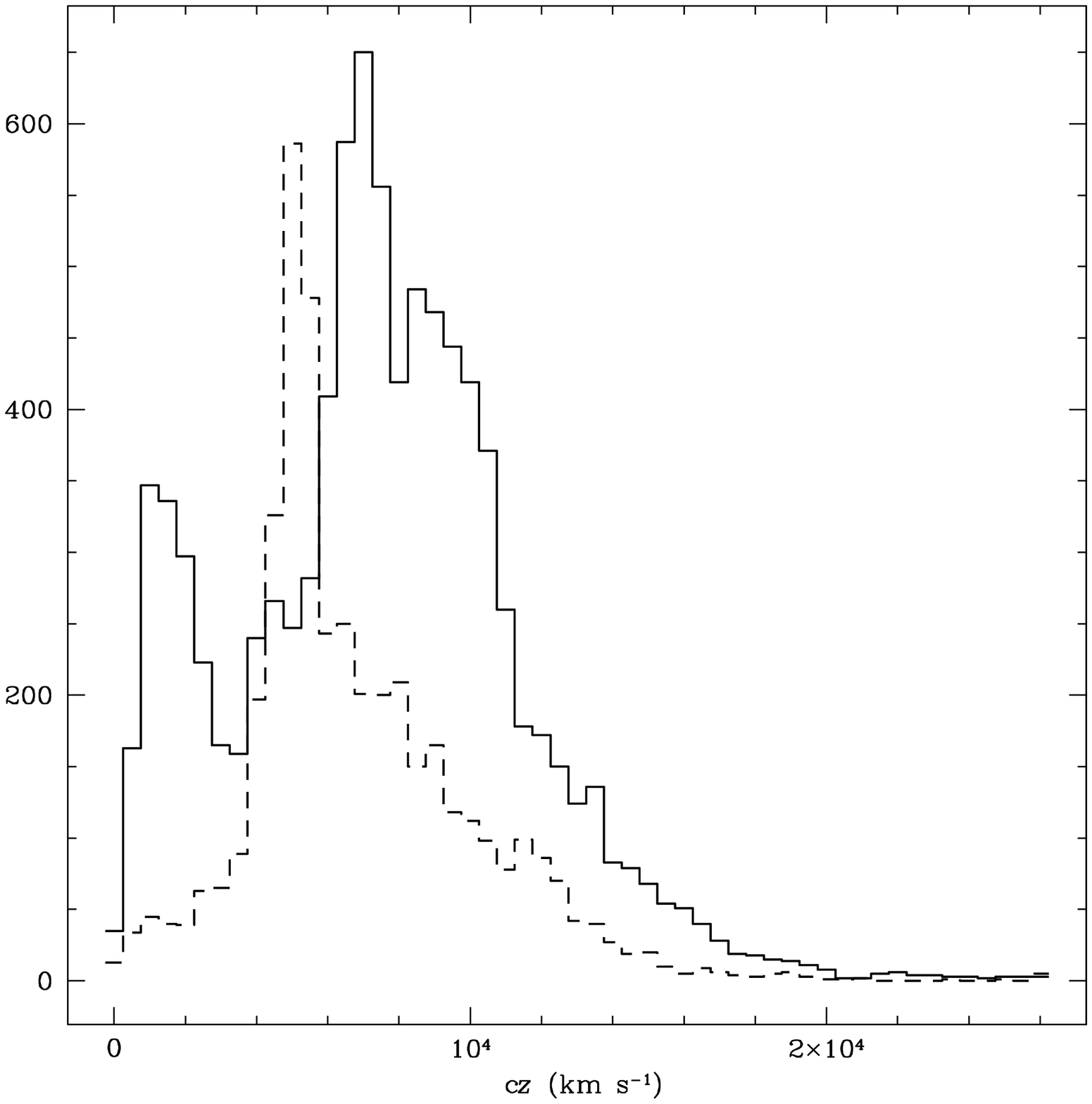} 
\figcaption{ 
The number of galaxies $N({\rm c}z)$ per redshift bin for the North (solid)
and South (dashed) galactic hemispheres. Note the prominence of the
Great Wall structure at c$z = 7,000-10,000$ \kms, and of the Virgo cluster
at c$z \sim 1,000$\kms.  Perseus-Pisces is also noticeable 
in the South at c$z \sim 5,000$ \kms. The differences between North 
and South reflect the inhomogeneities on the scale of these large surveys.}
\end{figure}

\newpage
\clearpage
\begin{deluxetable}{llrrccclcrlcl}
\scriptsize
\tablecaption{Sample UZC Catalog}
\tablewidth{0pt}
\tablehead{
\colhead{RA (J2000) DEC} 
& \colhead{$B$}& \colhead{c$z$} & \colhead{$\Delta$c$z$}
& \colhead{T}& \colhead{U} &\colhead{N}
& \colhead{ZNCAT Name}&\colhead{Ref.} &\colhead{Ref.} & \colhead{Other Name} 
& UZC & NED
\nl
\colhead{$hhmmss.s$\ $ddmmss$}
& \colhead{mag} & \colhead{\kms} & \colhead{\kms} 
& \colhead{} & \colhead{} & \colhead{} 
& \colhead{}& \colhead{code}& \colhead{} & \colhead{}
& Mult. & Mult.
}
\startdata
\ 130007.3+084155& 15.2&  13893&  53& B& 0&  0& 125736+08580& Z&     & 12576+0858    &  & T\nl
\ 130003.5+265353& 14.7&   5898&  70& -& 0&  0& 125736+27100&  & 0000& N4892         &  &  \nl
\ 130002.1+332613& 15.2&   7253&  35& E& 0&  0& 125742+33420& Z& 2700& 12577+3342    &  &  \nl
\ 130004.7+275914& 15.1&   6412&  25& A& 0&  1& 125739+28153& F&     & N4886         & *&  \nl
\ 130010.5+122900& 13.3&   1516&  71& A& 0&  0& 125742+12450& Z& 2700& N4880         &  &  \nl
\ 125956.4+471221& 15.5&   8696&  46& A& 0&  0& 125742+47280& Z&     & N4901         &  &  \nl
\ 130007.8+275838& 13.0&   6502&  48& A& 0&  2& 125742+28150& Z&     & N4889         & *&  \nl
\ 130013.9+284943& 15.4&   7365&  24& A& 2&  0& 125748+29060& F&     & 12578+2906    &  &  \nl
\ 130017.5+275720& 15.0&   6848&  15& -& 0&  1& 125753+28135&  & 3606& N4898W        & *& P\nl
\ 130017.8+281210& 14.3&   8430&  48& A& 0&  0& 125753+28281& Z&     & N4895         &  &  \nl
\ 130016.1+361515& 15.5&   8297&  47& B& 0&  0& 125754+36310& Z& 2700& I4028         &  &  \nl
\ 130024.9+134013& 15.4&   2030&  40& E& 1&  0& 125754+13570& Z& 0625& 12579+1357    &  &  \nl
\ 130022.0+280250& 15.5&   8153&  55& -& 0&  0& 125757+28189&  & 1823& I4026         &  &  \nl
\ 130025.6+285206& 15.4&   6812&  59& -& 0&  0& 125800+29080&  & 4200& I4032         &  &  \nl
\ 130033.2+100748& 15.1&   7169&  42& B& 0&  0& 125800+10240& Z&     & 12580+1024    &  &  \nl
\ 130029.2+264031& 15.2&   7208& 100& -& 0&  0& 125800+26560&  & 1502& 12580+2656    &  &  \nl
\ 130039.1+023000& 12.8&    936&  36& E& 0&  0& 125806+02460& Z& 0611& N4900         &  &  \nl
\ 130030.8+282047& 15.1&   5998&  28& A& 0&  0& 125806+28370& F&     & N4896         &  &  \nl
\ 130033.7+273816& 15.4&   7495&  15& E& 0&  0& 125806+27550& F&     & 12581+2755    &  &  \nl
\ 130037.7+280329& 15.1&   7840&   5& -& 0&  0& 125813+28196&  & 9000& I4040         &  &  \nl
\ 130039.7+275528& 15.2&   7505&  15& -& 0&  1& 125815+28116&  & 3606& N4906         & *&  \nl
\ 130039.6+290110& 14.6&   7289&   3& -& 0&  0& 125812+29170&  & 0661& I 842         &  &  \nl
\ 130042.6+275819& 15.5&   6255&  70& -& 0&  1& 125818+28142&  & 0000& I4042         & *&  \nl
\ 130047.6+154217& 15.5&   6503&  49& B& 0&  0& 125818+15590& Z&     & 12583+1559    &  &  \nl
\ 130054.5-012048& 15.2&   6813&  42& B& 0&  0& 125818-01050& Z& 2774& 12583-0105    &  &  \nl
\ 125956.2+734133& 14.7&   1665&   5& -& 1&  0& 125818+73580&  & 2218& 12583+7358    &  &  \nl
\ 130042.7+362044& 15.3&   8311&  52& A& 0&  0& 125824+36370& Z& 2700& I4049         &  &  \nl
\ 130042.9+371855& 12.7&   4660&  43& A& 0&  0& 125824+37350& Z& 2700& N4914         &  &  \nl
\ 130048.5+280528& 15.1&   6855&  15& -& 0&  1& 125824+28215&  & 3606& I4045         & *&  \nl
\ 130048.7+280931& 14.6&   5879&  10& -& 0&  0& 125824+28255&  & 9000& N4907         &  &  \nl
\ 130050.0+272420& 15.5&   6629&  36& A& 0&  0& 125824+27400& F&     & 12584+2740    &  &  \nl
\ 130059.0-000145& 13.2&   1153&  44& B& 1&  0& 125824+00140& Z& 2700& N4904         &  &  \nl
\ 130051.4+280235& 14.9&   8841&  70& -& 0&  2& 125827+28186&  & 3900& N4908         & *&  \nl
\ 130052.0+282159& 14.9&   7665&  23& A& 2&  0& 125830+28380& F&     & 12585+2838    &  &  \nl
\ 130103.6-015712& 15.3&   1419&  10& -& 0&  0& 125830-01420&  & 3000& 12585-0142    &  & P\nl
\ 130054.6+280026& 14.8&   5050&  30& A& 0&  1& 125829+28165& F&     & I4051         & *&  \nl
\ 130055.9+274728& 13.7&   7939&  51& A& 0&  0& 125830+28030& Z&     & N4911         &  &  \nl
\ 130055.6+471320& 15.0&   7381&  58& A& 0&  0& 125842+47300& Z& 2700& N4917         &  &  \nl
\ 130029.9+701155& 15.3&       &    & -& 2&  0& 125848+70280&  &     & 12588+7028    &  &  \nl
\ 130106.7+395029& 15.0&  10620&  46& A& 0&  0& 125848+40070& Z& 2700& 12588+4007    &  &  \nl
\ 130122.4+071907& 15.5&  13576&  59& A& 0&  0& 125848+07350& Z& 2774& 12588+0735    &  &  \nl
\ 130117.5+274834& 14.9&   7344&  31& A& 2&  0& 125854+28040& F&     & N4919         &  &  \nl
\ 130124.3+291832& 14.2&   7187&  24& A& 2&  0& 125900+29350& F&     & N4922A        &  & P\nl
\ 130125.1+284038& 15.3&   8775&  82& B& 0&  0& 125906+28570& F&     & 12591+2857    &  &  \nl
\ 130126.1+275309& 13.7&   5480&  31& A& 0&  1& 125900+28080& F&     & N4921         & *&  \nl
\ 130116.6+480337& 15.1&   8975&  14& E& 3&  1& 125900+48190& F&     & 12590+4819    & *& P\nl
\ 130118.3+480332& 15.1&   8888&  14& E& 3&  1& 125900+48190& F&     & 12590+4819    & *& P\nl
\ 130131.7+275052& 14.7&   5469&  31& A& 2&  1& 125906+28060& F&     & N4923         & *&  \nl
\ 130141.5+044050& 15.4&  11458&  47& B& 0&  0& 125912+04570& Z& 2774& 12592+0457    &  &  \nl
\ 130133.6+290750& 14.8&   7387&  25& -& 0&  0& 125912+29240&  & 2315& I 843         &  &  \nl
\enddata             
\end{deluxetable}

\newpage
\clearpage
\begin{deluxetable}{cp{12cm}}
\tablecaption{Velocity Sources for the UZC}
\tablewidth{0pt}
\tablenum{2}
\tablehead{
\colhead{ID} & \colhead{Bibliographic Reference}
\nl
}
\startdata
 0000 & de Vaucouleurs, G., de Vaucouleurs, A. \& Corwin, H. 1976, \par         
        The Second Reference Catalogue of Bright Galaxies \par                  
        University of Texas Press, Austin, Texas (RC2) \nl                      
 0001 & Corwin, H. G. Jr. \& Emerson, D. 1982, MNRAS  200, 621. \nl             
 0002 & de Vaucouleurs, G. et al. Third Reference Catalogue of Bright           
        Galaxies,\par                                                           
        Springer-Verlag, New York (RC3) \nl                                     
 0004 & de Vaucouleurs, G., de Vaucouleurs, A. \& Nieto, J. L. 1979, AJ 84,     
        1811 \nl                                                                
 0005 & Kelton, P. 1980, AJ  85, 89 \nl                                         
 0100 & Sandage, A. \& Tammann, G. 1981, The Revised Shapley-Ames Catalog \par  
        Carnegie Institution of Washington \nl                                  
 0104 & Sandage, A. 1976, PASP  88, 367 \nl                                     
 0105 & Sandage, A. 1978, AJ  83, 904, 1467 \nl                                 
 0109 & Binggeli, B., Sandage, A. \& Tammann, G. 1985, AJ  90, 1681 \nl         
 0200 & Fisher, J. R.  \& Tully, R. B.  1981, ApJS  47, 139 \nl                 
 0302 & Huchra, J. \& Sargent, W. L. W. 1973, ApJ  186, 433 \nl                 
 0400 & Rubin, V. et al. 1976, AJ  81, 687\nl                                   
 0402 & Rubin, V. et al. 1976, AJ  81, 719 \nl                                  
 0501 & Arakelyan, M., Dibai, E. \& Esipov, V. 1975, Astrofizika  11, 15 \nl    
 0502 & Arakelyan, M., Dibai, E. \& Esipov, V. 1975, Astrofizika  11, 377 \nl   
 0503 & Arakelyan, M., Dibai, E. \& Esipov, V. 1976, Astrofizika  12, 195 \nl   
 0504 & Arakelyan, M., Dibai, E. \& Esipov, V. 1976, Astrofizika  12, 685 \nl   
 0505 & Arkhipova, V. \& Esipov, V. 1979, Soviet Astron. Lett.  5, 140 \nl      
 0506 & Arkhipova, V., Esipov, V. \& Savel'eva, M. 1976, Soviet Astron. 20,     
        521 \nl                                                                 
 0508 & Denisyuk, E. \& Lipovetskii, V. 1977, Soviet Astron. Lett.  3, 3 \nl    
 0509 & Denisyuk, E., Lipovetskii, V. \& Afanasiev, V. 1976, Astrofizika 12,    
        665 \nl                                                                 
 0510 & Dibai, E., Doroshenko, V. \& Terebizh, V. 1976, Astrofizika 12, 689 \nl 
 0511 & Doroshenko, V. \& Terebizh, V. 1975, Astrofizika  11, 631 \nl           
 0512 & Kopylov, I. et al. 1976, Astrofizika  12, 189\nl                        
 0513 & Markaryan, B., Lipovetskii, V. \& Stepanyan, J. 1980, Astrofizika 16,   
        5 \nl                                                                   
 0514 & Markaryan, B., Lipovetskii, V. \& Stepanyan, J. 1980, Astrofizika 16,   
        609 \nl                                                                 
 0515 & Petrosyan, A., Saakyan, K. \& Khachikyan, E. 1979, Astrofizika  15,     
        373 \nl                                                                 
 0516 & Stephanyan, V. A. 1984, Astrofizika  21, 245 \nl                        
 0517 & Markaryan, B. E., Lipovetskii, V. A. \& Stepanyan, V. A. 1984,          
        Astrofizika 21, 419\nl                                                  
 0521 & Kazaryan, M. A. 1987, Astrofizika  27, 399 \nl                         
 0522 & Lipovetskii, V. A. et al. 1989, Astrofizika  31, 425 \nl                
 0523 & Kazaryan, M. \& Kazaryan, E. 1987, Astrofizika  26, 5 \nl               
 0525 & Denisyuk, E. \& Lipovetskii, V. 1974, Astrofizika  10, 315 \nl          
 0526 & Arakelyan, M., Dibai, E. \& Esipov, V. 1972, Astrofizika 8, 177 \nl     
 0527 & Arakelyan, M., Dibai, E. \& Esipov, V. 1972, Astrofizika 8, 329 \nl     
 0528 & Arakelyan, M., Dibai, E. \& Esipov, V. 1970, Astrofizika 6, 39 \nl      
 0529 & Arakelyan, M., Dibai, E. \& Esipov, V. 1972, Astrofizika 8, 33 \nl      
 0530 & Arakelyan, M., Dibai, E. \& Esipov, V. 1970, Astrofizika 7, 177 \nl     
 0600 & Giovanelli, R. \& Haynes, M. 1981, private communication \nl            
 0601 & Giovanelli, R. \& Haynes, M. 1985, ApJ  292, 404 \nl                    
 0602 & Williams, B. A. \& Lynch, J.R. 1991, AJ  101, 1969 \nl                  
 0604 & Chincarini, G. L., Giovanelli, R. \& Haynes, M. P. 1983, ApJ 269, 13\nl 
 0605 & Haynes, M. P. \& Giovanelli, R. 1991, AJ  102, 841 \nl                  
 0606 & Gordon, D. \& Gottesman, S. 1981, AJ  86, 161 \nl                       
 0607 & Freudling, W., Martel, H. \& Haynes, M. P. 1991, ApJ  377, 349 \nl      
 0609 & Krumm, N. \& Salpeter, E. 1979, ApJ  228, 64 \nl                        
 0610 & Salzer, J. J. 1992, AJ  103, 385 \nl                                    
 0611 & Krumm, N. \& Salpeter, E. 1980, AJ  85, 1312 \nl                        
 0612 & Helou, G., Salpeter, E. \& Krumm, N. 1979, ApJ  228, L1 \nl             
 0613 & Peterson, S. 1979, ApJS  40, 527 \nl                                    
 0614 & Olson, E. 1979, private communication \nl                               
 0615 & Schombert, J. M. et al. 1992, AJ  103, 1107 \nl                         
 0616 & Williams, B. \& Kerr, F. 1981, AJ  86, 953 \nl                          
 0617 & Haynes, M. \& Giovanelli, R. 1984, AJ  89, 758 \nl                      
 0618 & Fontanelli, P. 1984, A\&A  138, 85  \nl                                 
 0619 & Bothun, G. et al. 1985, AJ  90, 2487 \nl                                
 0620 & Giovanardi, C. \& Salpeter, E. 1985, ApJS  58, 623 \nl                  
 0621 & Scholl, J. \& Grayzeck, E. 1984, PASP  96, 216 \nl                      
 0622 & Giovanelli, R. \& Haynes, M. 1985, AJ  90, 2445. \nl                    
 0623 & Bicay, M. \& Giovanelli, R. 1986, AJ  91, 705 \& 732 \nl                
 0624 & Giovanelli, R. et al. 1986, AJ  92, 250 \nl                             
 0625 & Hoffman, G. L. et al. 1987, ApJS  63, 247 \nl                           
 0626 & Gavazzi, G. 1987, ApJ  320, 96 \nl                                      
 0627 & Bicay, M. \& Giovanelli, R. 1987, AJ  93, 1326 \nl                      
 0628 & Jackson, J. et al. 1987, AJ  93, 531 \nl                                
 0631 & Haynes, M. et al. 1988, AJ   95, 607 \nl                                
 0633 & Freudling, W., Haynes, M. \& Giovanelli, R. 1988, AJ  96, 1791 \nl      
 0634 & Hoffman, G. et al. 1989, ApJS  69, 65 \nl                               
 0636 & Sulentic, J. \& Arp. H. 1983, AJ  88, 489 \nl                           
 0637 & Eder, J. et al. 1989, ApJ  340, 29 \nl                                  
 0639 & Schneider, S. et al. 1990, ApJS  72, 245 \nl                            
 0642 & Giovanelli, R. \& Haynes, M. 1989, AJ  90, 633 \nl                      
 0643 & Lewis, B. M., Helou, G. \& Salpeter, E. E. 1985, ApJS  59, 161\nl       
 0644 & Lu, N. Y. et al. 1990, ApJ 357, 388 \nl                                 
 0645 & Mould, J. R. et al. 1993, ApJ 409, 14\nl                                
 0647 & Freudling, W., Haynes, M. P. \& Giovanelli, R. 1992, ApJS 79, 157 \nl   
 0649 & Wegner, G., Haynes, M. P. \& Giovanelli, R. 1993, AJ 105, 1251 \nl      
 0650 & Giovanelli, R. \& Haynes, M. P. 1993, AJ 105, 1271 \nl                  
 0651 & Lu, N. Y. et al. 1993, ApJS 88, 383 \nl                                 
 0653 & Hoffman, G. L., Lewis, B. M. \& Salpeter, E. E. 1995, ApJ 441, 28  \nl  
 0654 & Pantoja, C. A. et al. 1997, ApJ 113, 905 \nl                            
 0656 & Giovanelli, R., Avera, E. \& Karachentsev, I. D. 1997, AJ 114, 122\nl   
 0658 & Nordgren, T. E. et al. 1998, ApJS 115, 43 \nl                           
 0660 & Binggeli, B., Popescu, C. C. \& Tammann, G. A. 1993, A\&AS 98, 275\nl   
 0661 & Haynes, M. P. et al. 1997, AJ 113, 1197 \nl                             
 0662 & Scodeggio, M. et al. 1995, ApJ 444, 41 \nl                              
 0802 & Huchtmeier, W. \& Bohnenstengel, H. 1975, A\&A  44, 479 \nl             
 0803 & Huchtmeier, W., Tammann, G. \& Wendker, H. 1976, A\&A  46, 381 \nl      
 0805 & Richter, O.-G. \& Huchtmeier, W. 1982, A\&A  109, 155 \nl               
 0809 & Huchtmeier, W. K. 1997, A\&A 319, 401 \nl                               
 0810 & Huchtmeier, W. K. \& Skillman, E. D. 1998, A\&AS 127, 269 \nl           
 0811 & Huchtmeier, W. K., Hopp, U. \& Kuhn, B. 1997, A\&A 319, 67 \nl          
 0900 & Knapp, G. 1978, private communication \nl                               
 0902 & Giovanelli, R. \& Haynes, M. 1982, AJ  87, 1355 \nl                     
 0906 & Romanishin, W. 1980, private communication \nl                          
 0908 & Shostak, G. 1978, A\&A  68, 321\nl                                      
 0909 & Thonnard, N., Rubin, V., Ford, K. \& Roberts, M. 1978, AJ  83, 1564 \nl 
 0910 & Richter, O.-G. \& Huchtmeier, W.K. 1987, A\&AS  68, 427 \nl             
 0912 & Tifft, W. G. \& Cocke, W. J. 1988, ApJS  67, 1  \nl                     
 0913 & Haynes, M.P. \& Giovanelli, R. 1991, ApJS  77, 331 \nl                  
 0923 & Dickey, J. M. 1997, AJ 113, 1939 \nl                                    
 0930 & Schneider, S. E. et al. 1992, ApJS 81, 5 \nl                            
 1001 & Bohuski, T., Fairall, A. \& Weedman, D. 1978, ApJ  221, 776 \nl         
 1009 & Peterson, C. 1978, PASP  90, 10 \nl                                     
 1023 & Vader, J. P. \& Chaboyer, B. 1992, PASP, 104, 57 \nl                    
 1032 & Kirhakos, S. D. \& Steiner, J. E. 1990, AJ 99, 1722 \nl                 
 1105 & Hintzen, P. 1980, AJ  85, 626 \nl                                       
 1108 & Kirshner, R., Oemler, A. \& Schechter, P. 1978, AJ  83, 1549 \nl        
 1109 & Kirshner, R. 1977, ApJ  212, 319 \nl                                    
 1111 & Peterson, B. M. 1978, ApJ  223, 740 \nl                                 
 1114 & Ulrich, M. H. 1978, ApJ  221, 422 \nl                                   
 1118 & Keel, W. 1985, AJ  90, 2207 \nl                                         
 1119 & Schweizer, L. 1987, ApJ  64, 411 \nl                                    
 1143 & Willmer, C. N. A. et al. 1996, ApJS 104, 199\nl                         
 1149 & Slinglend, K. et al. 1998, ApJS 115, 1 \nl                              
 1154 & Owen, F. N., Ledlow, M. J. \& Keel, W. C. 1995, AJ 109, 14 \nl          
 1155 & Caldwell, N. \& Rose, J. A. 1997, AJ 113, 492 \nl                       
 1203 & Gonz\'alez-Serrano, J. I. \& Valentijn, E. A. 1991, A\&A  242, 334 \nl  
 1207 & Huchtmeier, W. \& Richter, O.-G. 1984, A\&A  149, 118\nl                
 1231 & Richter, O.-G. 1987, A\&A  67, 261\nl                                   
 1235 & Schulte-Ladbeck, R. 1988, PASP  100, 785 \nl                            
 1243 & De Grijp, M., Miley, G. \& Lub, J. 1985, Nature  314, 240 \nl           
 1244 & Moorwood, A., V\'eron-Cetty, M.-P. \& Glass, I. 1987, A\&A  184, 63\nl  
 1261 & Stickel, M. et al. 1991, ApJ  374, 431 \nl                              
 1300 & Davis, M. 1979, private communication \nl                               
 1301 & Schild, R. \& Davis, M. 1979, AJ  84, 311 \nl                           
 1502 & Gregory, S. 1976, ApJ  199, 1 \nl                                       
 1503 & Gregory, S. \& Thompson, L. 1978, ApJ  222, 784 \nl                     
 1507 & Hintzen, P. et al. 1983, AJ  87, 1656 \nl                               
 1514 & Tarenghi, M. et al. 1979, ApJ  234, 793 \nl                             
 1516 & Thompson, L., Welker, W. \& Gregory, S. 1978, PASP  90, 644 \nl         
 1520 & Tifft, W. \& Gregory, S. 1976, ApJ  205, 696 \nl                        
 1521 & Tifft, W. \& Gregory, S. 1979, ApJ  231, 23 \nl                         
 1524 & Tifft, W. \& Gregory, S. 1988, AJ  95, 651 \nl                          
 1526 & Stocke, J. T. et al. 1987, ApJ 315, L11  \nl                            
 1528 & Kailey, W. \& Lebofsky, M. 1988, ApJ  326, 653 \nl                      
 1537 & Ledlow, M. J. et al. 1996, AJ 112, 388  \nl                             
 1538 & Pinkney, J. et al. 1993, ApJ 416, 36 \nl                                
 1602 & Chincarini, G. \& Rood, H. 1972, AJ  77, 448 \nl                        
 1604 & Chincarini, G. \& Rood, H. 1976, ApJ  206, 30 \nl                       
 1605 & Chincarini, G. \& Rood, H. 1977, ApJ  214, 351 \nl                      
 1606 & Dickel, J. \& Rood, H. 1978, ApJ  223, 391 \nl                          
 1609 & Rood, H. \& Dickel, J. 1976, ApJ  205, 346 \nl                          
 1733 & Caganoff, S., Bicknell, G. V. \& Carter, D. 1985, PASA 6, 151 \nl       
 1802 & Afanasiev, V. et al. 1980, A\&A  91, 302 \nl                            
 1803 & Arkhipova, V. et al. 1981, Astrofizika  17, 240 \nl                     
 1804 & Karachentsev, I. 1980, ApJS  44, 137 \nl                                
 1805 & Karachentsev, I. \& Karachentseva, V. 1981, Astrofizika  17, 5 \nl      
 1806 & Karachentsev, I. \& Karachentseva, V. 1982, Soviet Astron. Lett. 8,     
        104 \nl                                                                 
 1810 & Karachentsev, I., Pronik, V. \& Chuvaev, K. 1975, A\&A  41, 375 \nl     
 1811 & Karachentsev, I., Pronik, V. \& Chuvaev, K. 1976, A\&A  51, 185 \nl     
 1814 & Kostyuk, I., Karachentsev, I. \& Kopylov, A. 1981, Soviet Astron.       
        Lett.  7, 148 \nl                                                       
 1815 & Kostyuk, I. \& Kopylov, A. 1982, Soviet Astron. Lett. 8, 280 \nl        
 1818 & Markaryan, B. E., Lipovetskii, V. A. \& Stepanyan, D. A. 1984,          
        Astrofizika  20, 213\nl                                                 
 1819 & Lipovetskii, V. A. \& Stepanyan, J. A. 1986, \par                       
        FBS The First Byurakan Sky Survey, Catalogue of Markarian Galaxies,\par 
        Report from the Special Astrophysical Observatory\nl                    
 1820 & Kostyuk, I. P. 1975, Soobshch. Spets. Astrofiz. Obs. Akad. Nauk SSR     
        13, 45 \nl                                                              
 1822 & Karachentsev, I. D. 1983, Soviet Astron. Lett.  9, 36 \nl               
 1823 & Karachentsev, I. D. \& Kopylov, A.I. 1990, MNRAS  243, 390 \nl          
 1824 & Karachentsev, I. D. 1981, Soviet Astron. Lett.  7, 1 \nl                
 1901 & Ulrich, M.-H. 1975, A\&A  40, 337 \nl                                   
 1908 & Wills, D. \& Wills, B. 1982, AJ  87, 252 \nl                            
 2000 & Zwicky, F. 1971, Catalogue of Selected Compact Galaxies and \par        
        of Post-Eruptive Galaxies, Zwicky, G\"umligen, Switzerland \nl          
 2101 & Eastmond, T. \& Abell, G. 1978, PASP  90, 367 \nl                       
 2102 & Faber, S. \& Dressler, A. 1977, AJ  82, 187 \nl                         
 2109 & Shuder, J. \& Osterbrock, D. 1981, ApJ  250, 55 \nl                     
 2118 & Willick, J. A., Brodie, J. P. \& Bowyer, S. 1990, ApJ  355, 393 \nl     
 2134 & Dey, A., Strauss, M. \& Huchra, J. 1989, AJ  99, 463\nl                 
 2139 & Martel, A. \& Osterbrock, D. E. 1994, AJ 107, 1283  \nl                 
 2200 & Bottinelli, L., Gouguenheim, L. \& Paturel, G. 1981, A\&AS  44, 217 \nl 
 2201 & Balkowski, C., Chamaraux, P. \& Weliachew, L. 1978, A\&A  69, 263 \nl   
 2204 & Bottinelli, L., Gouguenheim, L. \& Paturel, G. 1980, A\&A  88, 32 \nl   
 2205 & Bottinelli, L., Gouguenheim, L. \& Paturel, G. 1982, A\&AS  50, 101 \nl 
 2208 & Chamaraux, P. 1987, 1988 private communication \nl                      
 2210 & Hamabe, M. et al. 1988, PASP  40, 47 \nl                                
 2212 & Bottinelli, L. et al. 1990. A\&AS 82, 391 \nl                           
 2214 & Garcia, A. M. et al. 1994, A\&AS 107, 265 \nl                           
 2216 & Tamazian, V. S., Theureau, G. \& Coudreau-Durand, N. 1997, A\&AS 126,   
        471 \nl                                                                 
 2218 & Theureau, G. et al. 1998, A\&AS 130, 333 \nl                            
 2306 & Leech, K. et al. 1988, MNRAS 231, 977\nl                                
 2308 & Lawrence, A. et al. 1989, MNRAS 240, 329 \nl                            
 2315 & Lucey, J. R. et al. 1991, MNRAS 253, 584 \nl                            
 2324 & Smith, R. J. et al. 1997, MNRAS 291, 461\nl                             
 2334 & Clements, D. L. et al. 1996, MNRAS 279, 459 \nl                         
 2407 & Karachentsev, I., Sargent, W. L. W. \& Zimmerman, B. 1979,              
        Astrofizika  15, 25 \nl                                                 
 2408 & Kent, S. 1978, Ph.D. Thesis, Caltech \nl                                
 2411 & Kunth, D. \& Sargent, W. L. W. 1979, A\&AS  36, 259 \nl                 
 2412 & Kunth, D. \& Sargent, W. L. W. 1979, A\&A  76, 50\nl                    
 2417 & Sulentic, J. 1980, A\&A  88, 94 \nl                                     
 2422 & Kirshner, R. et al. 1987, ApJ  314, 493 \nl                             
 2430 & Sargent, W. L. 1968, ApJ  153, L135 \nl                                 
 2443 & Sargent, W. L. W. 1970, ApJ 160, 405 \nl                                
 2449 & Small, T. A., Sargent, W. L. W. \& Hamilton, D. 1997, ApJS  111, 1 \nl  
 2452 & Dale, D. A. et al. 1998, AJ 115, 418\nl                                 
 2454 & Dale, D. A. et al. 1997, AJ 114, 455 \nl                                
 2457 & Schweizer, F. 1996, AJ 111, 109 \nl                                     
 2600 & Allen, R. \& Shostak, G. 1979, A\&AS  35, 163 \nl                       
 2602 & Shostak, G. \& Allen, R. 1980, A\&A  81, 167 \nl                        
 2605 & van Driel, W. \& van Woerden, H. 1989, A\&A  225, 317\nl                
 2606 & Bosma, A., van der Hulst, J. M. \& Sullivan, W. T. 1977, A\&A 57,       
        373\nl                                                                  
 2607 & Bosma, A., van der Hulst, J. \& Athanassoula, E. 1988, A\&A 198, 100\nl 
 2608 & van Driel, W., Davies, R. \& Appleton, P. 1988, A\&A 199, 41\nl         
 2609 & Kamphuis, J., Sijbring, D. \& van Albada, T. 1996, A\&AS 116, 15\nl     
 2700&  Huchra, J. et al. 1983, ApJS  52, 89 \nl                                
 2706 & Elvis, M. et al. 1981, ApJ  246, 20 \nl                                 
 2707 & Huchra, J., Wyatt, W. \& Davis, M. 1982, AJ  87, 1628 \nl               
 2709 & Schwartz, D. et al. 1980, ApJ  238, L53\nl                              
 2710 & Shectman, S., Stefanick, R. \& Latham, D. 1983, AJ  88, 477 \nl         
 2712 & White, S. et al. 1982, MNRAS  203, 701\nl                               
 2714 & Bothun, G. et al. 1983, ApJ  268, 47\nl                                 
 2715 & Geller, M. et al. 1984, AJ  89, 319\nl                                  
 2716 & Huchra, J. \& Brodie, J. 1984, ApJ  280, 547\nl                         
 2717 & Beers, T. et al. 1984, ApJ  283, 33\nl                                  
 2718 & Huchra, J. et al. 1985, AJ 90, 691\nl                                   
 2720 & Foltz, C. \& Chaffee, F. 1987, AJ  93, 529 \nl                          
 2722 & Fabricant, D. et al. 1986, ApJ  308, 530 \nl                            
 2724 & Smith, B. et al. 1987, ApJ  318, 161\nl                                 
 2725 & Chapman, G. N. F., Geller, M. J. \& Huchra, J., 1987, AJ 94, 571\nl     
 2728 & Chapman, G. N. F., Geller, M. J. \& Huchra, J., 1987, AJ 95, 999 \nl    
 2729 & Strauss, M. \& Huchra, J. 1988, AJ  95, 1602 \nl                        
 2731 & Zabludoff, A., Huchra, J. \& Geller, M., 1990, ApJS  74, 1 \nl          
 2732 & Michel, A. \& Huchra, J. 1988, PASP  100, 1423 \nl                      
 2733 & Ostriker, E. C. et al. 1988, AJ  96, 1775 \nl                           
 2734 & Fabricant, D., Kent, S. \& Kurtz, J. 1989, ApJ  336, 77\nl              
 2736 & Huchra, J. et al. 1990, ApJS  72, 433\nl                                
 2740 & Rawlings, S., Eales, S. \& Warren, S. 1990, MNRAS 243, 14\nl            
 2743 & Beers, T. et al. 1991, AJ 102, 1581\nl                                  
 2748 & Hickson, P. et al. 1992, ApJ 399, 353 \nl                               
 2754 & Strauss, M. et al. 1992, ApJS 83, 29\nl                                 
 2755 & Beers, T. C. et al. 1992, ApJ 400, 410\nl                               
 2760 & Zabludoff, A. I. et al. 1993, AJ 106, 1273 \nl                          
 2761 & Dell'Antonio, I., Geller, M. \& Fabricant, D. 1994, AJ 107, 427 \nl     
 2762 & Huchra, J. P. et al. 1993, AJ 105, 1637\nl                              
 2764 & Beers, T. et al. 1995, AJ 109, 874\nl                                   
 2765 & Fisher, K. et al. 1995, ApJS 100, 69\nl                                 
 2766 & Marzke, R. O., Huchra, J. P. \& Geller, M. J. 1996, AJ 112, 1803 \nl    
 2767 & Ramella, M. et al. 1995, AJ 109, 1458 \nl                               
 2769 & Barmby, P. \& Huchra, J. P. 1998, AJ 115, 6  \nl                        
 2773 & Hughes, J. P. \& Birkinshaw, M. 1998, ApJ 497, 645 \nl                  
 2774 & Grogin, N. A., Geller, M. J. \& Huchra, J. P. 1998 ApJS \nl             
 2777 & Barton, E. J., de Carvalho, R. R. \& Geller, M. J. 1998,                
        astro-ph/9806397 \nl                                                    
 2802 & Staveley-Smith, L., Davies, R. D. \& Kinman, T. D. 1992, MNRAS 258,     
        334\nl                                                                  
 2812 & Hurley, S. 1988, private communication \nl                              
 2813 & Axon, D. et al. 1988, MNRAS 231, 1077\nl                                
 3000 & Thuan, T. \& Seitzer, P. 1979, ApJ  231, 327\nl                         
 3114 & Metcalfe, N. 1989, MNRAS 236, 207 \nl                                   
 3118 & Fairall, A. P. et al. 1992, AJ 103, 11 \nl                              
 3200 & Bothun, G. et al. 1985, ApJS 57, 423 \nl                                
 3200 & Bothun, G. 1981, Ph.D. Thesis, University of Washington \nl             
 3201 & Heckman, T., Balick, B. \& Sullivan, W. 1978, ApJ 224, 745\nl           
 3203 & Sullivan, W. et al. 1981, AJ 86, 919\nl                                 
 3300 & Rood, H. 1981, private communication \nl                                
 3501 & Barbieri, C. et al. 1979, A\&AS  37, 559\nl                             
 3502 & Merighi, R. et al. 1991, A\&AS  89, 225 \nl                             
 3505 & Malamuth, E. M. \& Kriss, G. A. 1986, ApJ 308, 10\nl                    
 3506 & Quintana, H. et al. 1985, AJ  90, 410\nl                                
 3507 & Vennik, Y. \& Kaazik, A. 1985, Astrofizika  23, 213 \nl                 
 3508 & Proust, D. et al. 1987 A\&AS 67, 57\nl                                  
 3509 & Focardi, P., Marano, B. \& Vettolani, G. 1986, A\&A 161, 217\nl         
 3517 & Kulessa, A.S. \& Lyndon-Bell, D. 1992, MNRAS 255, 105 \nl               
 3520 & Stickel, M. \& Kuhr, H. 1993, A\&AS 100, 395\nl                         
 3527 & Davoust, E. \& Consid\`ere, S. 1995, A\&AS 110, 19 \nl                  
 3528 & Di Nella, H. et al. 1995, A\&AS 113, 151 \nl                            
 3542 & Impey, C. D. et al. 1996, ApJS 105, 209 \nl                             
 3601 & Barbon, R. et al. 1982, A\&AS  49, 73 \nl                               
 3604 & Davis, M., Sargent, W. \& Tonry, J. 1989, private communication \nl   
 3606 & Davies, R. et al. 1987, ApJS  64, 581\nl                                
 3608 & Augarde, R. et al. 1987, A\&A  185, 4 \nl                               
 3611 & Dressler, A. \& Shectman, S.A. 1988, AJ 95, 284 \nl                     
 3624 & Zaritsky, D. et al. 1997, ApJ 478, 39 \nl                               
 3700 & Palumbo, G., Tanzella-Nitti, G. \& Vettolani, G. 1983, \par             
        Catalogue of Radial Velocities of Galaxies,                             
        (New York:Gordon \& Breach)\nl                                          
 3800 & Huchtmeier, W. K. \& Richter, O.-G. 1989, \par                          
        A General Catalog of HI Observations of Galaxies, (Berlin:\par          
        Springer-Verlag) \nl                                                    
 3900 & da Costa, L. N. et al. 1984, AJ 89, 1310\nl                             
 3905 & da Costa, L. N. et al. 1988 ApJ 327, 544 \nl                            
 4001 & Bothun, G. et al. 1989, ApJS  70, 271 \nl                               
 4002 & Salzer, J., MacAlpine, G. \& Boroson, T. 1989, ApJS,  70, 447 \nl       
 4005 & Fabricant, D. et al. 1993, AJ 105, 788\nl                               
 4007 & Mohr, J. J., Geller, M. J. \& Wegner, G. 1996, AJ 112, 1816\nl          
 4105 & Hill, G. et al. 1988, AJ  95, 1031\nl                                   
 4200 & Wegner, G. et al. 1990 AJ  100, 1405 \nl                                
 4201 & Wegner, G., Haynes, M. P. \& Giovanelli, R. 1993, AJ 105, 1251 \nl      
 4300 & Davis, M. \& Strauss, M. 1990, private communication \nl                
 4302 & Strauss, M. A. et al. 1992, ApJS 83, 29 \nl                             
 4701 & Lawrence, A. et al. 1994 MNRAS, 266, 41\nl                              
 4730 & Paturel, G. et al. 1989, A\&AS 80, 299  \nl                             
 4900 & Bothun, G. \& Mould, J. 1992, private communication \nl                 
 5011 & Hardin, D. IAU Circ 6629 \nl                                            
 5502 & Bottinelli, L. et al. 1993, A\&AS 102, 57 \nl                           
 5507 & Lanzetta, K. M., Webb, J. K. \& Barcons, X. 1996, ApJ 456, L17 \nl      
 5510 & Veron-Cetty, M.-P. \& V\'eron, P. 1993, A\&AS 100, 521 \nl              
 5511 & Augarde, R. et al. 1994, A\&AS 104, 259 \nl                             
 5519 & Ebeling, H., Mendes de Oliveira, C. \& White, D. A. 1995, MNRAS 277,    
        1006 \nl                                                                
 5521 & Jorgensen, I., Franx, M. \& Kjaergaard, P. 1995, MNRAS 276, 1341 \nl    
 5523 & Vettolani, G. et al. 1998, astro-ph/9805195 \nl                         
 5600 & The NED Team, ned.ipac.caltech.edu \nl                                  
 9000 & Casoli, F., Dickey, J., Kaz\`es, I, Boselli, A., Gavazzi, G. and        
        Jore, K 1996, A\&A\ S 116, 193 \nl                                      
\enddata
\end{deluxetable}

\end{document}